\documentclass[twocolumn,times]{aastex62}

\usepackage{color}

\newcommand {\valrmin} {2.0}
\newcommand {\valrmax} {21.1}
\newcommand {\Nsample} {34}

\newcommand {\valrmaxcomb} {39.5}
\newcommand {\Nsamplecomb} {46}

\newcommand {\valgamma} {3.53}
\newcommand {\valegamma} {0.01}

\newcommand {\valbeta} {0.46}
\newcommand {\valepbeta} {0.15}
\newcommand {\valembeta} {0.19}

\newcommand {\valconc} {16.8}
\newcommand {\valalpha} {0.26}

\newcommand {\valf} {1.01}
\newcommand {\valepf} {0.17}
\newcommand {\valemf} {0.16}

\newcommand {\valMrmax} {0.21}
\newcommand {\valepMrmax} {0.04}
\newcommand {\valemMrmax} {0.03}

\newcommand {\valvrmax} {206}
\newcommand {\valepvrmax} {19}
\newcommand {\valemvrmax} {16}

\newcommand {\valMvir} {1.28}
\newcommand {\valepMvir} {0.97}
\newcommand {\valemMvir} {0.48}

\newcommand {\valbetacomb} {0.52}
\newcommand {\valepbetacomb} {0.11}
\newcommand {\valembetacomb} {0.14}

\newcommand {\valfcomb} {1.00}
\newcommand {\valepfcomb} {0.14}
\newcommand {\valemfcomb} {0.13}

\newcommand {\valMrmaxcomb} {0.42}
\newcommand {\valepMrmaxcomb} {0.07}
\newcommand {\valemMrmaxcomb} {0.06}

\newcommand {\valvrmaxcomb} {214}
\newcommand {\valepvrmaxcomb} {17}
\newcommand {\valemvrmaxcomb} {15}

\newcommand {\valMvircomb} {1.54}
\newcommand {\valepMvircomb} {0.75}
\newcommand {\valemMvircomb} {0.44}

\newcommand {\kms} {km\,s$^{-1}$}
\newcommand {\Msun} {M$_\odot$}

\newcommand {\vLOS} {v_\mathrm{LOS}}
\newcommand {\rmin} {r_\mathrm{min}}
\newcommand {\rmax} {r_\mathrm{max}}
\newcommand {\rvir} {r_\mathrm{virial}}
\newcommand {\Mvir} {M_\mathrm{virial}}
\newcommand {\rscale} {r_\mathrm{scale}}
\newcommand {\vcirc} {v_\mathrm{circ}}
\newcommand {\rbreak} {r_\mathrm{break}}
\newcommand {\gammain} {\gamma_\mathrm{in}}
\newcommand {\gammaout} {\gamma_\mathrm{out}}
\newcommand {\Mrmax} {M \left( < \rmax \right)}
\newcommand {\Mgridrmax} {M_\mathrm{grid} \left( < \rmax \right)}
\newcommand {\Mgridvir} {M_\mathrm{grid,virial}}
\newcommand {\Mtme} {M_\mathrm{TME}}
\newcommand {\Mtrue} {M_\mathrm{true}}
\newcommand {\vrmax} {\vcirc \left( \rmax \right)}
\newcommand {\vtan} {v_\mathrm{tan}}

\newcommand {\rrange} {$\mathrm{\valrmin} \le r \le \mathrm{\valrmax}$~kpc}

\newcommand {\rrangecomb} {$\mathrm{\valrmin} \le r \le \mathrm{\valrmaxcomb}$~kpc}

\graphicspath{{./}{figures/}}

\shorttitle{Intermediate-Mass Milky Way from Gaia DR2}
\shortauthors{Watkins et al.}

\begin{document}

\title{Evidence for an Intermediate-Mass Milky Way from \textit{Gaia} DR2 Halo Globular Cluster Motions}

\correspondingauthor{Laura Watkins}
\email{lwatkins@stsci.edu}

\author[0000-0002-1343-134X]{Laura L. Watkins}
\affiliation{Space Telescope Science Institute, 3700 San Martin Drive, Baltimore MD 21218, USA}

\author[0000-0001-7827-7825]{Roeland P. van der Marel}
\affiliation{Space Telescope Science Institute, 3700 San Martin Drive, Baltimore MD 21218, USA}
\affiliation{Center for Astrophysical Sciences, Department of Physics \& Astronomy, Johns Hopkins University, Baltimore MD 21218, USA}

\author[0000-0001-8368-0221]{Sangmo Tony Sohn}
\affiliation{Space Telescope Science Institute, 3700 San Martin Drive, Baltimore MD 21218, USA}

\author[0000-0002-5981-7360]{N. Wyn Evans}
\affiliation{Institute of Astronomy, University of Cambridge, Madingley Road, Cambridge, CB3 0HA, UK}

\begin{abstract}
We estimate the mass of the Milky Way (MW) within \valrmax~kpc using the kinematics of halo globular clusters (GCs) determined by \textit{Gaia}. The second \textit{Gaia} data release (DR2) contained a catalogue of absolute proper motions (PMs) for a set of Galactic globular clusters and satellite galaxies measured using \textit{Gaia} DR2 data. We select from the catalogue only halo GCs, identifying a total of \Nsample\ GCs spanning \rrange, and use their 3D kinematics to estimate the anisotropy over this range to be $\beta = \valbeta^{+\valepbeta}_{-\valembeta}$, in good agreement, though slightly lower than, a recent estimate for a sample of halo GCs using \textit{HST} PM measurements further out in the halo. We then use the \textit{Gaia} kinematics to estimate the mass of the MW inside the outermost GC to be $M(<\valrmax \; \mathrm{kpc}) = \valMrmax^{+\valepMrmax}_{-\valemMrmax} \times 10^{12} \mathrm{M_\odot}$, which corresponds to a circular velocity at $\rmax$ of $\vcirc (\valrmax \; \mathrm{kpc}) = \valvrmax^{+\valepvrmax}_{-\valemvrmax} \mathrm{km\,s^{-1}}.$ The implied virial mass is $\Mvir = \valMvir^{+\valepMvir}_{-\valemMvir} \times 10^{12} \mathrm{M_\odot}$. The error bars encompass the uncertainties on the anisotropy and on the density profile of the MW dark halo, and the scatter inherent in the mass estimator we use. We get improved estimates when we combine the \textit{Gaia} and \textit{HST} samples to provide kinematics for  \Nsamplecomb\ GCs out to $\valrmaxcomb$~kpc: $\beta = \valbetacomb^{+\valepbetacomb}_{-\valembetacomb}$, $M(<\valrmaxcomb \; \mathrm{kpc}) = \valMrmaxcomb^{+\valepMrmaxcomb}_{-\valemMrmaxcomb} \times 10^{12} \mathrm{M_\odot}$, and $\Mvir = \valMvircomb^{+\valepMvircomb}_{-\valemMvircomb} \times 10^{12} \mathrm{M_\odot}$. We show that these results are robust to potential substructure in the halo GC distribution. While a wide range of MW virial masses have been advocated in the literature, from below $10^{12}$~\Msun\ to above $2 \times 10^{12}$~\Msun, these new data imply that an intermediate mass is most likely.
\end{abstract}

\keywords{dark matter -- Galaxy: fundamental parameters -- Galaxy: halo -- Galaxy: kinematics and dynamics -- Galaxy: structure -- globular clusters: general}

\section{Introduction}
\label{section:introduction}

The mass of the Milky Way (MW) is one of its most fundamental parameters, and yet, despite decades of intense effort, our best estimates are significantly scattered, with some estimates agreeing very well, and others differing by more than their uncertainties \citep[see][for a thorough review]{BlandHawthorn2016}. These estimates are very sensitive to assumptions made in the modelling, including, but not limited to, which of the MW's satellites are bound or unbound and for how long they have been bound, the velocity anisotropy of the MW halo and of its satellite system, the shape of the MW halo, and the particular method used for the analysis. Estimates typically range from as low as $\sim 0.5 \times 10^{12}$~\Msun\ \citep[e.g.][radial anisotropy with Leo I unbound]{Watkins2010} to as high as $2-3 \times 10^{12}$\Msun\ from abundance-matching studies \citep[e.g.][]{BoylanKolchin2010}, the timing argument \citep{Li2008, vanderMarel2012a}, or studies of tracers \citep[e.g.][tangential anisotropy with Leo I bound]{Watkins2010}.

Accurate determination of the mass profile of the MW has implications for our understanding of the dynamical history of the Local Group \citep[both past evolution and future interactions, e.g.][]{vanderMarel2012a, vanderMarel2012b} and the MW's satellite population, particularly the Sagittarius dSph and its impressive tidal stream \citep{Fardal2019}, and the Magellanic Clouds \citep{Kallivayalil2013}.

Furthermore, the mass of a galaxy and its distribution (or shape) are intrinsically linked to the formation and growth of structure in the Universe \citep{Conselice2014}, so accurately determining these parameters for the MW will give us a clearer understanding of where our Galaxy sits in a cosmological context \citep[for an excellent review see][]{Freeman2002}. In particular, we can know whether the MW is typical or atypical, and thus, how much of what we learn about the MW can be safely assumed for other galaxies as well.

The MW is composed of a central nucleus that harbours a supermassive black hole (SMBH) at its heart, a bulge, a disk, and a halo \citep{Ivezic2012, BlandHawthorn2016}. The first three components are all primarily baryonic in nature, and while many of their properties remain topics of some debate, their masses are reasonably well determined. The final component, the halo, is dominated by dark matter (DM) -- only a few percent of the mass of the halo is baryonic \citep{Helmi2008}, the exact percentage depends on the unknown total mass of DM in the halo -- and it is our inability to see DM directly that gives rise to our present uncertainty in the mass.

As the majority of mass in the MW is `invisible', we cannot measure it directly, instead we can infer its presence by its influence on its surroundings. Typically, this is the purview of dynamical studies. Any mass distribution gives rise to a gravitational potential that causes objects to move: by studying measurements of the motions of the objects, we can work backwards to recover the underlying gravitational potential and, thus, the mass distribution.

There are some mass-estimation methods, notably the timing argument and abundance-matching studies, that estimate the total mass of a system. However, most dynamical methods work by using tracer objects to probe the properties of the whole system, and can only estimate the mass over the range for which tracer data are available. Thus different families of tracers provide crucial information at different points depending on the range they cover. This is particularly crucial in the MW where globular clusters (GCs) tend to probe the inner regions of the halo, while dwarf spheroidal (dSph) satellite galaxies offer better coverage further out~\citep[e.g.,][]{Wilkinson1999,Watkins2010,Pa18}. The modelling of multiple stellar streams may provide a promising alternative~\citep[e.g.,][]{Gibbons2014,Sa17}, though six-dimensional phase space data is only available for the GD-1 stream at present~\citep{Koposov2010,Bowden2015}.

One key problem with mass estimation via kinematics is that we need to know the total velocity of each tracer, but we are seldom fortunate enough to have all 3 components of motion for a large sample of tracers. Typically, we only have line-of-sight (LOS) velocities. This is especially troublesome for studies of the MW as the Sun is very close to the Galactic Centre, and so for most objects LOS velocities predominantly probe only one component of the motion (the Galactocentric radial direction) and offer little information about the Galactocentric tangential motions of the tracers\footnote{The degree to which the LOS and Galactocentric radial velocities are similar depends on the geometry of the system, specifically the position of the object relative to both the Sun and the Galactic Centre, and there are a few objects with more favourable geometry to give some insight on the tangential motions. However, anisotropy measurements rely on averaging over many objects, so favourable geometry for a few objects provides limited benefit overall.}. With only LOS velocities, the masses we estimate depend very strongly on what assumptions we make for the tangential motions: the well-known mass-anisotropy degeneracy \citep[e.g.][]{Binney2008}.

Some methods attempt to overcome the lack of 3D velocity information: \citet{Eadie2015} introduced a Galactic Mass Estimator that includes unknown velocity components as free parameters in their models. However, the best constraints on mass will come from having complete phase space information and, with proper motions (PMs), we are able to break this degeneracy. Firstly, we can make a direct estimate of the velocity anisotropy, and secondly, we can correctly include the total velocity of the tracers in our mass calculations instead of having to make assumptions.

Absolute PMs have been measured from the ground for a number of GCs \citep[e.g.][and other papers in the series]{CasettiDinescu2013}, although typically this is only possible with sufficient accuracy for objects within $\sim$10~kpc, and even then ground-based measurements often suffer from a number of systematic effects. Space offers a more stable environment for astrometry, so thanks to its excellent precision and long time baseline, the \textit{Hubble Space Telescope} (\textit{HST}) has proved extremely valuable for providing absolute PMs for dSphs \citep[e.g.][]{Piatek2016, Sohn2017} and GCs \citep[e.g.][]{Anderson2003, Kalirai2007}. The recent study by \citet{Sohn2018} that measured PMs using \textit{HST} for 20 outer halo GCs in the MW represents the largest sample of absolute PMs measured to date in a single study.

The \textit{Gaia} mission's \citep{Gaia} first data release \citep{GaiaDR1} contained proper motions for $\sim$2 million stars in the Tycho-Gaia Astrometric Solution (TGAS) catalogue \citep{Michalik2015}, which used \textit{Tycho2} \citep{Hog2000} measurements to provide a first epoch of data and \textit{Gaia} data for the second, and has been used already for multiple PM studies of objects in the MW, including for a handful of Galactic GCs. \citet{Watkins2017} identified member stars for 5 GCs in the TGAS data and used the stars to estimate the absolute PMs of their host clusters; comparing these \textit{Gaia} PMs with previous estimates, they found excellent agreement with previous \textit{HST} measurements, but some differences to previous ground-based values due to systematics inherent in such measurements. \citet{Massari2017} used archival \textit{HST} data combined with \textit{Gaia} DR1 data to estimate the PM of GC NGC\,2419.

However, the second \textit{Gaia} data release \citep{GaiaDR2} has greatly expanded our view of the local Universe. This data release provides PMs for billions of stars and has made it possible to measure absolute PMs for 75 Galactic GCs out to a Galactocentric distance of $\sim$21~kpc, along with 9 classical dSphs, a single ultrafaint dwarf, and both the Large and Small Magellanic Clouds \citep{GaiaGCs}. This is by far the largest catalogue of GC and dSph PMs to date. Combined with position and LOS velocity information from previous studies, these measurements have enabled analysis of the orbits of these objects.

In this paper, we use these motions to provide new mass estimates for the MW. In \autoref{section:data}, we describe the Gaia catalogues, calculate the Galactocentric motions of the objects, and describe which objects we select for our analysis; in \autoref{section:mwmass}, we estimate the mass of the MW; in \autoref{section:discussion}, we compare our results with previous estimates; and in \autoref{section:conclusions}, we summarise our findings.

\section{Overview of Data}
\label{section:data}

\subsection{\textit{Gaia} Halo Cluster Sample}
\label{section:Gaiaclusters}

\citet{GaiaGCs} used the the Gaia DR2 Catalogue \citep{GaiaDR2} to identify member stars for a number of MW GCs and dSphs based on their positions, photometry, and PMs, from which their absolute PMs were calculated\footnote{Catalogues from \url{https://www.astro.rug.nl/~ahelmi/research/dr2-dggc/}}. Combined with distances and line-of-sight velocities, it was then possible to calculate the orbit of each object within the Galaxy. The orbits derived do depend on certain assumptions made for the potential of the Galaxy, which, as we have discussed, is still somewhat uncertain. To mitigate the effects of this uncertainty, orbits were calculated in 3 different potentials that span a range of possible MW shapes and masses. We will use both the absolute PMs and the orbital properties here.

The first step is to calculate Galactocentric positions and motions from the observed heliocentric values. We take right ascensions, declinations, and the PMs in these coordinates, along with the full covariance matrix for the PM uncertainties from \citet{GaiaGCs}.\footnote{We have not included the \textit{Gaia} systematic errors of $\sim$0.035~mas/yr as, at the distances we are considering, these errors translate to a few km/s and will be very much smaller than the velocity dispersion of the halo.} Distances and LOS velocities, we take from \citep[][2010 edition]{Harris1996}, which are primarily determined photometrically; these distances generally agree with kinematical distances determined from \textit{HST} PM studies \citep{Watkins2015}. We assume distance uncertainties $\Delta D = 0.023 D$, which is equivalent to uncertainties on the distance moduli of 0.05~mag, which is typical for GC distance uncertainties \citep{Dotter2010}. There are a few GCs in the Harris catalogue for which there is an LOS velocity measurement but no uncertainty listed. For the GCs with uncertainties, the average is $\overline{\Delta \vLOS} = 0.06 \vLOS$, which we thus adopt for the remaining GCs. We assume a distance from the Sun to the Galactic Centre of $R_\odot = 8.29 \pm 0.16$~kpc and a circular velocity at the solar radius of $V_\odot = 239 \pm 5$~\kms\ \citep[both][]{McMillan2011}. For the solar peculiar velocity relative to the Local Standard of Rest, we assume $\mathbf{V_\mathrm{pec}} = (11.10 \pm 1.23, 12.24 \pm 2.05, 7.25 \pm 0.62)$~\kms\ \citep{Schoenrich2010}. We will use these solar parameters throughout the paper.

We calculate the positions and velocities of the GCs in a spherical coordinate system (radius $r$, latitude $\theta$, and longitude $\phi$) centred on the Galactic Centre. We use Monte Carlo sampling, using 1000 samples and assuming Gaussian uncertainties, to propagate all of the observational uncertainties -- this includes the full covariance matrix for the PMs, uncertainties of the distances and LOS velocities, and the all uncertainties on the position and velocity of the Sun. Although the initial distributions are assumed Gaussian, the resulting distributions of Galactocentric properties may not be, so we take medians and 15.9 and 84.1 percentiles of the distributions as the best estimate and uncertainties.\footnote{For Gaussian distributions, the 15.9 and 84.1 percentiles enclose the 1-$\sigma$ confidence interval. The posterior distributions we derive thoroughout are generally not Gaussian, so that these uncertainties should be interpreted as actual percentiles ranges, with any analogy to Gaussian errors only being approximate at best.} These Galactocentric positions and velocities are provided in \autoref{appendix:galactocentric}.

One of the main contributors to the uncertainty in the mass of the MW is the paucity of tracer objects. The dSphs are limited in number and appear to have been accreted in groups~\citep{GaiaGCs}, which is an extremely interesting result but makes mass modelling tricky as certain key assumptions are thus invalidated. For the GCs, on the other hand, the improvements in PM accuracy offered by \textit{Gaia} over \textit{HST} (where such data exists) are modest with only 22 months of \textit{Gaia} data. Where \textit{Gaia} can play a key role here is to provide Galactocentric motions for many more clusters than were previously available, greatly increasing our sample size. As such, we choose to concentrate on the GC sample henceforth.

To probe the anisotropy and mass of the Galactic halo, we require a sample of halo clusters, free of disk and bulge contaminants. \citet{Zinn1993} showed that the disk and bulge clusters separate cleanly from the halo clusters in metallicity. We use the same cut and keep only clusters with $[\mathrm{Fe/H}] \le -0.8$~dex \citep[metallicities from][2010 edition]{Harris1996} so as to have a pure halo sample. Furthermore, the mass estimators we will use assume that the potential is scale-free over the region of interest, so we wish to limit ourselves to clusters for which this assumption reasonably holds, that is we do not wish to include clusters that spend most of their time in the innermost regions of the Galaxy where the disk is a significant contributor and the potential is non-spherical. We use the orbital parameters from \citet{GaiaGCs} to extract only GCs with apocentres\footnote{Orbits were calculate in 3 different potentials; we insisted that at least 2 of the 3 apocentre estimates must pass the cut.} $r_\mathrm{apo} \ge 6$~kpc\footnote{As we will see later, this value is equivalent to two disk scale lengths for our adopted disk model. We experimented with different halo samples, and found that our results were not sensitive to the particular choice of apocentre cut that we used.}; this leaves us with \Nsample\ GCs that span a radial range \rrange.

\subsection{\textit{HST} Halo Cluster Sample}
\label{section:HSTclusters}

Recently, \citet{Sohn2018} presented \textit{HST} PMs for 20 halo GCs that extend further out into the halo than the \textit{Gaia} cluster sample. Four of these are in common with the \citet{GaiaGCs} sample: NGC\,2298, NGC\,5024, NGC\,5053, and NGC\,5466. The \textit{HST} estimates are in good agreement with the \textit{Gaia} estimates for three of the four clusters, which is reassuring news for both catalogues. The measurements for the fourth cluster differ by $\sim$48~\kms\, but this is still well below the velocity dispersion of the halo (see \autoref{table:results_kinematics}). This also indicates that we can confidently combine the catalogues and increase the size of our sample and its range. This is an improvement on both analyses as the \textit{Gaia} catalogue probes further in and the \textit{HST} catalogue probes further out, but they also have a substantial region of overlap to serve as a solid anchor and consistency check.

We follow our approach from \citet{Sohn2018} and exclude NGC\,2419 as its distance is much greater than the rest of the sample\footnote{As mass estimates depend strongly on $r$, a single cluster far out in the halo can unduly bias any mass estimates.}, and three of the four clusters associated with the Sagittarius dSph as they represent a group, not a well mixed population (and including all four can again lead to biases). We also choose to use the \textit{Gaia} values for the clusters in common. Overall, this gives us an extra 12 clusters in our sample, bringing the total to \Nsamplecomb, and increasing the radial range of the sample to $\valrmin \le r \le \valrmaxcomb$~kpc.

In what follows, we will provide analysis using only the \textit{Gaia} cluster sample (Sample A), and using the combined \textit{Gaia} and \textit{HST} cluster samples (Sample B).

\begin{figure}
    \centering
    \includegraphics[width=0.99\linewidth]{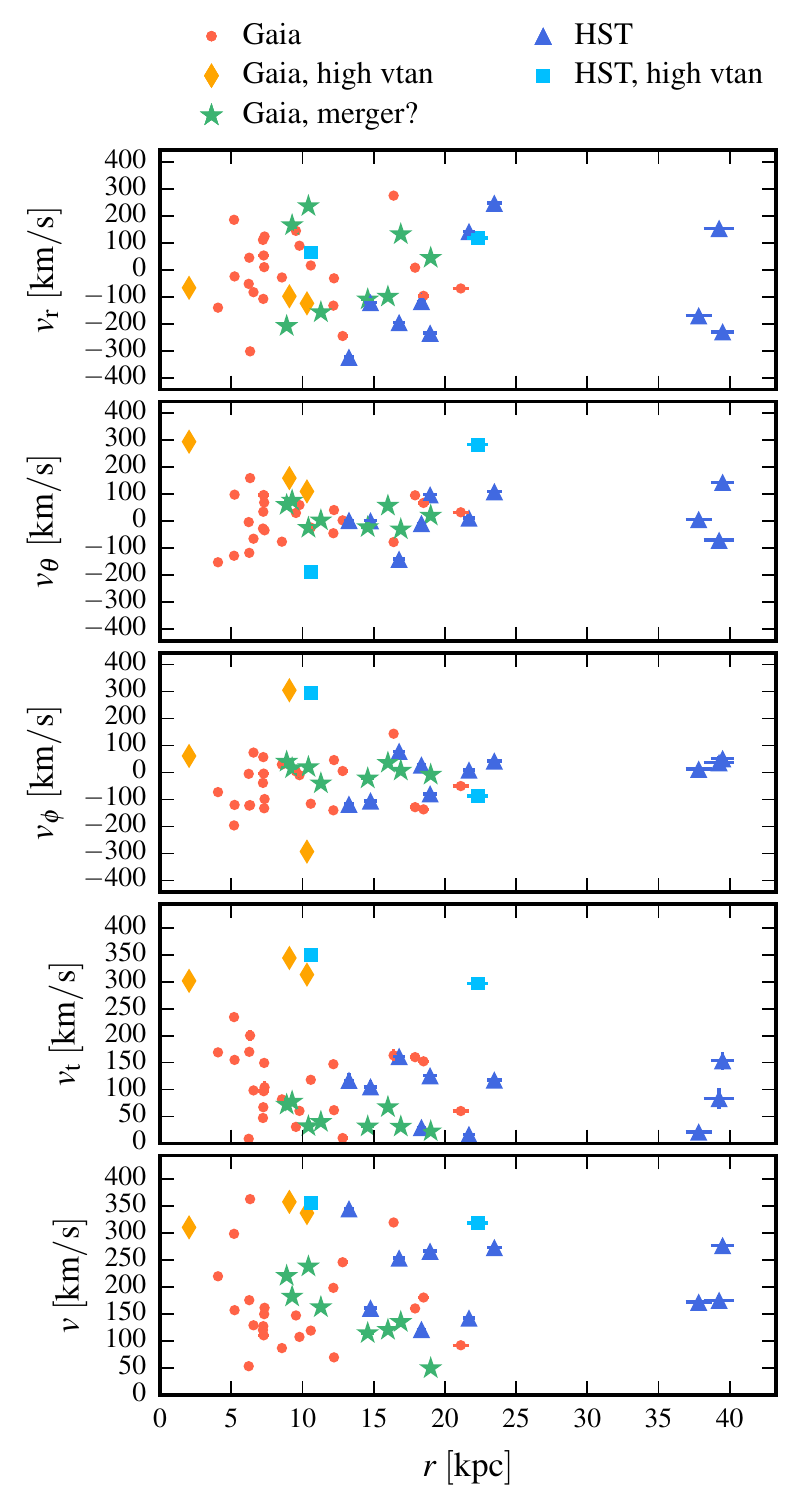}
    \caption{Galactocentric velocity distributions as a function of Galactocentric distance of the halo GCs, from top to bottom: Galactocentric radial velocities, Galactocentric latitudinal velocities, Galactocentric longitudinal velocities, Galactocentric tangential velocities, Galactocentric total velocities. In all panels, red circles, orange diamonds, and green stars show the \textit{Gaia} measurements and the cyan squares and blue triangles show the \textit{HST} measurements. Orange diamonds (\textit{Gaia}) and cyan squares (\textit{HST}) highlight the GCs with $\vtan \ge 250$~\kms, and the green stars highlight the clusters tentatively identified as part of a recent merger -- these subsamples are discussed in later sections. Uncertainties are shown in the figure but are too small to be visible in most cases.}
    \label{figure:data}
\end{figure}

\autoref{figure:data} shows the distribution of the Galactocentric velocities of the halo GC sample as a function of Galactocentric distance. The upper three panels show the radial $v_\mathrm{r}$, latitudinal $v_\theta$, and longitudinal $v_\phi$ velocity components. The next panel shows the tangential velocity where $v_\mathrm{t} = \sqrt{v_\theta^2 + v_\phi^2}$, and the bottom panel shows the total velocity $v$. The red circles, orange diamonds, and green stars show the halo GCs in the \textit{Gaia} DR2 sample. The blue triangles and cyan squares show the halo GCs from \textit{HST} measurements. The different groupings identify various subsamples that we will use later to verify the robustness of results against substructure.

\section{Milky Way Mass}
\label{section:mwmass}

We can use the Galactocentric positions and motions of the clusters to estimate the mass of the Milky Way within the radius of the outermost cluster in our sample, which lies at $\rmax = \valrmax$~kpc. \citet{Watkins2010} introduced a family of simple, yet effective, tracer mass estimators (TMEs), which we will use here. Subsequent extensions and applications of this method were presented in \citet{Annibali2018} in a study of NGC\,4449, and in \citet{Sohn2018} in a similar study of MW halo GCs.

The estimators work with different types of distance and velocity data, depending on what is available. As we have full 6D phase-space information for our cluster sample, we are able to use the estimator that uses distances and total velocities, equation 24 of \citet{Watkins2010}. That is,
\begin{equation}
    \Mrmax_\mathrm{TME} = \frac{\alpha+\gamma-2\beta}{G \left( 3-2\beta \right)} \rmax^{1-\alpha} \left\langle v^2 r^\alpha \right\rangle .
	\label{eqn:TME}
\end{equation}
The estimators assume that the underlying potential is a power law with index $\alpha$ over the region of interest, that the tracer objects have a number density distribution that is a power law with index $\gamma$ over the region of interest, and that the velocity anisotropy of the tracer sample is a constant $\beta$ over the region of interest. Before we can proceed, we need to estimate $\alpha$, $\beta$, and $\gamma$.

\subsection{Anisotropy}
\label{section:anisotropy}

In \autoref{section:data}, we calculated Galactocentric motions $v_\mathrm{j}$ of all the clusters in our sample in a spherical coordinate system $(j,k = \{ r, \theta, \phi \})$, along with a full covariance matrix for their uncertainties  $\delta_\mathrm{j}$ and the correlations between the uncertainties $\rho_\mathrm{jk}$. That is, for each cluster $i$, we have a velocities
\begin{equation}
	\mathbf{v}_\mathrm{i} = \left( v_\mathrm{r}, v_\theta, v_\phi \right)_\mathrm{i}
\end{equation}
with uncertainties
\begin{equation}
	\mathbf{S}_\mathrm{i} = \left( \begin{array}{ccc}
    	\delta_\mathrm{r}^2 & \rho_\mathrm{r \theta} \delta_\mathrm{r} \delta_\theta & \rho_\mathrm{r \phi} \delta_\mathrm{r} \delta_\phi \\
        \rho_\mathrm{r \theta} \delta_\mathrm{r} \delta_\theta & \delta_\theta^2 & \rho_{\theta \phi} \delta_\theta \delta_\phi \\
        \rho_\mathrm{r \phi} \delta_\mathrm{r} \delta_\phi & \rho_{\theta \phi} \delta_\theta \delta_\phi & \delta_\phi^2
    \end{array} \right)_\mathrm{i} .
\end{equation}

We assume that the cluster population has a mean velocity
\begin{equation}
	\overline{\mathbf{v}} = (\overline{v_\mathrm{r}}, \overline{v_\theta}, \overline{v_\phi})
\end{equation}
and covariance
\begin{equation}
	\mathbf{C} = \left( \begin{array}{ccc}
    	\sigma_\mathrm{r}^2 & 0 & 0 \\
        0 & \sigma_\theta^2 & 0 \\
        0 & 0 & \sigma_\phi^2
    \end{array} \right)
\end{equation}
where $(\overline{v_\mathrm{r}}, \overline{v_\theta}, \overline{v_\phi})$ are the mean velocities and $(\sigma_\mathrm{r}, \sigma_\theta, \sigma_\phi)$ the velocity dispersions for each coordinate. In setting the cross-terms of the covariance to zero, we have assumed that the axes of the velocity ellipsoid are aligned with the spherical coordinate system. We further assume that the velocity distributions are Gaussian. Then the likelihood of the observed measurements for mean $\overline{\mathbf{v}}$ and covariance $\mathbf{C}$ is,
\begin{equation}
	\mathcal{L} = \prod_i^N \frac{ \exp 
    	\left[ - \frac{1}{2}
        	\left( \mathbf{v}_\mathrm{i} - \overline{\mathbf{v}} \right)^{\mathrm{T}}
        	\left( \mathbf{C} + \mathbf{S}_\mathrm{i} \right)^{-1}
        	\left( \mathbf{v}_\mathrm{i} - \overline{\mathbf{v}} \right)
        \right] }
        { \sqrt{
        \left( 2 \mathrm{\pi} \right)^{3}
        \left| \left( \mathbf{C} + \mathbf{S}_\mathrm{i} \right) \right| } }.
\end{equation}

We use flat priors for the mean velocities in each coordinate, that is
\begin{equation}
	P(\overline{v}_\mathrm{j}) = 1 .
\end{equation}
We insist that the dispersions must be positive, but otherwise use a flat prior for positive dispersion values, that is
\begin{equation}
	P(\sigma_\mathrm{j}) = \left\{ \begin{array}{cc}
    	1 & \qquad \sigma_\mathrm{j} \ge 0 \\
        0 & \qquad \sigma_\mathrm{j} < 0
    \end{array} \right.
\end{equation}
Finally, the posterior is the product of the likelihood and the priors.

To estimate the means and dispersions that best describe the data, we use the affine-invariant Markov Chain Monte Carlo (MCMC) package \textsc{emcee} \citep{ForemanMackey2013} to find the region of parameter space where the posterior is maximised and to sample that region. We draw 10\,000 points from the final posterior distribution for our final sample. For each parameter, we adopt the median as the best estimate and use the 15.9 and 84.1 percentiles for the uncertainties.

\begin{figure}
    \centering
    \includegraphics[width=0.8\linewidth]{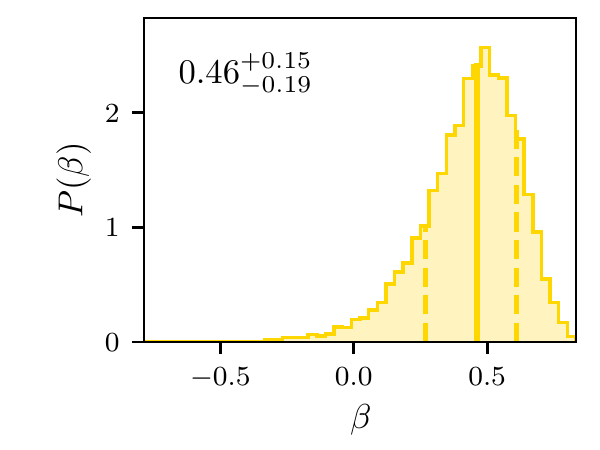}
    \caption{The posterior distribution of anisotropy $\beta$ for the \textit{Gaia} DR2 halo cluster sample.}
    \label{figure:anisotropy}
\end{figure}

From these, we are then able to estimate the anisotropy
\begin{equation}
    \beta = 1 - \frac{\sigma^2_\mathrm{\theta} + \sigma^2_\mathrm{\phi}}{2 \sigma^2_\mathrm{r}}
	\label{eqn:anisotropy}
\end{equation}
of the system. For each of the 10\,000 points in our final sample, we calculate the anisotropy $\beta$. The distribution of these anisotropy values is shown in \autoref{figure:anisotropy}, and has a median and 15.9- and 84.1-percentile uncertainties of $\beta = \valbeta^{+\valepbeta}_{-\valembeta}$. We discuss this value in the context of previous work in \autoref{section:discussion}.

\begin{deluxetable*}{cccccccc}
\tablecaption{Galactocentric mean velocities, dispersions, and anisotropies for the GC subsamples. \label{table:results_kinematics}}
\tablehead{
   \colhead{Sample} & 
   \colhead{$\overline{v_\mathrm{r}}$} & 
   \colhead{$\overline{v_\theta}$} &
   \colhead{$\overline{v_\phi}$} &
   \colhead{$\sigma_\mathrm{r}$} & 
   \colhead{$\sigma_\theta$} &
   \colhead{$\sigma_\phi$} &
   \colhead{$\beta$} \\
   \colhead{} & 
   \colhead{(\kms)} & 
   \colhead{(\kms)} & 
   \colhead{(\kms)} &
   \colhead{(\kms)} & 
   \colhead{(\kms)} & 
   \colhead{(\kms)}
}
\startdata
A\tablenotemark{a} & $-15.9^{+23.4}_{-24.0}$ & $ 22.1^{+15.6}_{-14.9}$ & $-29.8^{+18.6}_{-19.0}$ & $138.9^{+20.1}_{-15.8}$ & $ 91.3^{+13.0}_{-10.7}$ & $109.5^{+15.9}_{-12.4}$ & $0.46^{+0.15}_{-0.19}$ \\
B\tablenotemark{b} & $-26.7^{+23.2}_{-23.5}$ & $ 20.8^{+14.8}_{-14.5}$ & $-18.8^{+16.0}_{-16.3}$ & $153.3^{+18.0}_{-14.9}$ & $100.7^{+12.0}_{-10.3}$ & $110.4^{+13.8}_{-11.0}$ & $0.52^{+0.11}_{-0.14}$ \\
C\tablenotemark{c} & $-21.5^{+26.9}_{-26.2}$ & $ 22.2^{+21.0}_{-21.0}$ & $-41.0^{+24.0}_{-24.2}$ & $132.9^{+22.6}_{-16.7}$ & $103.7^{+17.2}_{-13.4}$ & $124.0^{+21.2}_{-15.5}$ & $0.24^{+0.23}_{-0.31}$ \\
D\tablenotemark{d} & $-27.1^{+23.9}_{-25.0}$ & $  7.9^{+12.1}_{-11.7}$ & $-27.5^{+11.9}_{-12.5}$ & $159.1^{+19.8}_{-16.4}$ & $ 76.4^{+ 9.8}_{- 8.3}$ & $ 78.4^{+10.0}_{- 8.4}$ & $0.76^{+0.06}_{-0.08}$ \\ 
\enddata
\tablenotetext{a}{\textit{Gaia} GCs.}
\tablenotetext{b}{\textit{Gaia} and \textit{HST} GCs.}
\tablenotetext{c}{\textit{Gaia} GCs with possible merger group removed.}
\tablenotetext{d}{\textit{Gaia} and \textit{HST} GCs with high $\vtan$ clusters removed.}
\end{deluxetable*}

\autoref{table:results_kinematics} summarises our fits to the velocity ellipsoid of the halo. We provide estimates of the mean velocities and velocity dispersions for each velocity component and the inferred anisotropy. The \textit{Gaia} sample is Sample A in the table. The other samples are described later in the text. We see that the \textit{Gaia} sample has a mean radial velocity consistent with 0 within its uncertainty, but shows hints of net tangential motion, which we will address later.

\subsection{Density}
\label{section:density}

To estimate the power-law index $\gamma$ of the halo cluster number density profile, we start with Galactocentric distances from the \citet[][2010 edition]{Harris1996} catalogue of 157 Galactic globular clusters. This catalogue contains both clusters that move on bulge-like and disk-like orbits and are found in the inner regions of the galaxy, and clusters that move on halo-like orbits and are found further out. All the clusters in our PM sample were deliberately chosen to be part of the halo cluster population, so that is the number density profile of interest for this analysis. We use a least-squares fitting algorithm to fit (with uncertainties) a broken power law to the data that has an index $\gammain$ in the inner regions and an index $\gammaout$ in the outer regions, with a break radius of $\rbreak$. We assume that $\gammain$ describes the density profile of the bulge and disk clusters and that $\gammaout$ describes the density profile of the halo clusters.

\begin{figure}
    \centering
    \includegraphics[width=0.9\linewidth]{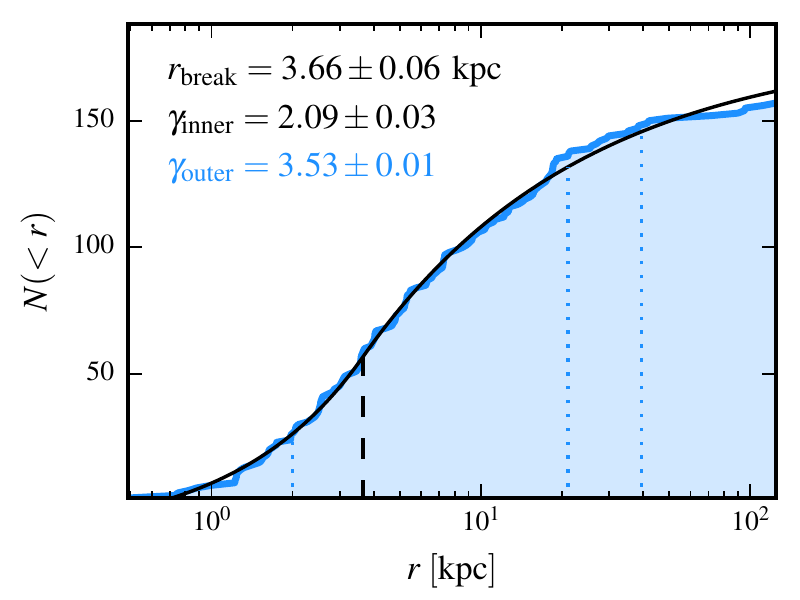}
    \caption{A power-law fit to the cumulative number profile of Milky Way globular clusters. To this we fit a broken power-law model that has an index $\gamma_\mathrm{in}$ in the inner regions, which we assume to be dominated by disk clusters, and an index $\gamma_\mathrm{out}$ in the outer regions, which we assume to be dominated by halo clusters, and is the value of interest for our analysis. The cumulative histogram in blue shows the data, the solid black line shows the best-fitting broken power law, and the dashed black line marks the break radius. The parameters for the best-fitting broken power law are shown in the top-left corner of the figure. The index for the outer power law is the value of interest and is highlighted in blue. The uncertainty on this value is the uncertainty in the fit, and does not account for uncertainty in the estimated positions of the Milky Way clusters or the functional form of the fit. For guidance, we also mark as blue dotted lines the radius of the innermost cluster in the \textit{Gaia} sample at $\valrmin$~kpc, the outermost clusters in the \textit{Gaia} sample at $\valrmax$~kpc, and the outermost cluster in the \textit{HST} sample at $\valrmaxcomb$~kpc.}
    \label{figure:density_double_powerlaw}
\end{figure}

\autoref{figure:density_double_powerlaw} shows the cumulative number density profile of all Milky Way globular clusters in blue. The solid black line is the best-fitting broken power law to the data, and the dashed black line marks the break radius at which the power-law index changes. The parameters of the fit are shown in the top-left corner, with uncertainties that are purely statistical and do not encompass systematic effects. As we assume that the outer profile describes the halo clusters, we adopt this value $\gamma = \valgamma \pm \valegamma$ for our mass analysis.\footnote{The fitted double power-law provides a better fit to the cumulative number profile inside $\sim$20~kpc than outside, as there are far fewer clusters at large radii to constrain the outer fits. Mass estimates correlate mildly with $\gamma$ such that a change in $\gamma$ of 0.2 will change the mass estimate inside $\valrmax$~kpc by $\sim$5\%, but we do not include this in our calculations.} This is the same value we adopted in a recent study of halo GCs with HST PMs by \citet{Sohn2018} and agrees well with previous studies \citep[eg.][]{Harris2001}.

\subsection{Potential}
\label{section:potential}

To estimate the power-law index $\alpha$ of the gravitational potential, we assume that the Milky Way is composed of a nucleus, bulge, disk, and halo. We adopt the nucleus, disk, and bulge prescriptions from \citet{PriceWhelan2017}, that is, we assume a Hernquist nucleus with mass $M_\mathrm{nucleus} = 1.71 \times 10^9$~\Msun and scale length $l_\mathrm{nucleus} = 0.07$~kpc, a Hernquist bulge with mass $M_\mathrm{bulge} = 5 \times 10^9$~\Msun and scale length $l_\mathrm{bulge} = 1$~kpc, and a Miyamoto-Nagai disk $M_\mathrm{disk} = 6.8 \times 10^{10}$~\Msun, scale length $l_\mathrm{disk} = 3$~kpc and scale height $h_\mathrm{disk} = 0.28$~kpc.

The shape and mass of the halo is uncertain -- indeed, this uncertainty is the primary motivation for the current analysis -- and so we choose to sample a range of possible halos to see what $\alpha$ values they imply. To do this, we assume that the halo is spherical and \citet[][NFW]{Navarro1996} in form, but with unknown virial radius $\rvir$ and a concentration $c$, which we will sample.\footnote{Our goal here is to approximate the slope of the potential over the region of interest. The precise functional form of the potential is unknown. We are assuming here that picking a single functional form and varying its parameters will give a similar distribution of $\alpha$ values as sampling a variety of functional forms. Especially as we insist that that circular velocity at the solar radius must be consistent with observations.}

The oft-cited study by \citet{Klypin2002} favours an NFW halo with virial mass $\Mvir = 1 \times 10^{12}$~\Msun\ and scale radius $\rscale = 21.5$~kpc (virial radius $\rvir = 258$~kpc and a concentration $c = 12$). More recent halo estimates have been less massive: \citet{Bovy2015} favours an NFW halo with virial mass $\Mvir = 0.8 \times 10^{12}$~\Msun\ and scale radius $\rscale = 16.0$~kpc (virial radius $\rvir = 245$~kpc and a concentration $c = 15.3$), and \citet{PriceWhelan2017} favours an NFW halo with virial mass $\Mvir = 0.54 \times 10^{12}$~\Msun\ and scale radius $\rscale = 15.62$~kpc (virial radius $\rvir = 214$~kpc and a concentration $c = 13.7$). However, other studies have found somewhat larger total masses for the Milky Way \citep[e.g.][]{Watkins2010} or concentrations \citep[e.g.][]{Deason2012} than these halo parameters would imply. As such, we choose to sample a range of virial radii $200 \le \rvir \le 400$~kpc, sampled at 1~kpc intervals, and a range of concentrations $8 \le c \le 20$, sampled at 0.1 intervals.

To further narrow down the list of allowed halos, we insist that the circular velocity at the solar radius must be consistent with observed values. Estimates for the circular velocity typically span $220-250$~\kms; for each halo, we estimate the circular velocity for the best-fitting power-law model at $R_\odot$ and reject halos with velocities outside of this range.

\begin{figure}
    \centering
    \includegraphics[width=0.9\linewidth]{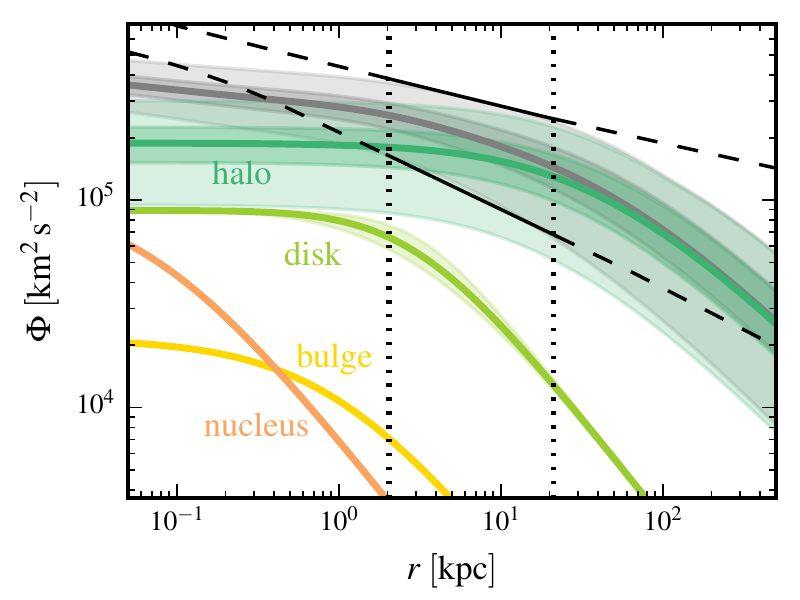}
    \caption{Range of power-law fits to the Milky Way potential. We assume that the Milky Way consists of a Hernquist nucleus (orange), a Hernquist bulge (yellow), a Miyamoto-Nagai disk (light green), and an NFW halo (dark green). Their sum (the total potential) is shown in grey. We assume that the nucleus, bulge, and disk components are fixed. The disk potential shows some slight broadening around 2 kpc as we have plotted potential as a function of spherical radius but the disk is not spherical. We sample a range of NFW halo parameters so, for the halo and total potentials, the solid lines show the median profiles, the dark shaded regions show the range between the 25th and 75th percentiles of the profiles, and the light shaded regions show the range of profiles for all halos sampled. The dotted lines show the region \rrange\ spanned by our halo cluster sample. The solid black lines show the extent of the best-fitting power laws in this region, and the dashed lines show the extent of the best-fitting power laws extended outside of the region of interest.}
    \label{figure:potential_powerlaw}
\end{figure}

For each halo model, we calculate the total potential profile from the nucleus, bulge, disk, halo components, and used a least-squares fitting algorithm to fit a power-law across the range spanned by our cluster sample \rrange. The index of the power-law fit $\alpha$ is the quantity we require for our models. \autoref{figure:potential_powerlaw} shows the range of potentials sampled and the range of the power law fits to those halos.

\begin{figure}
    \centering
    \includegraphics[width=\linewidth]{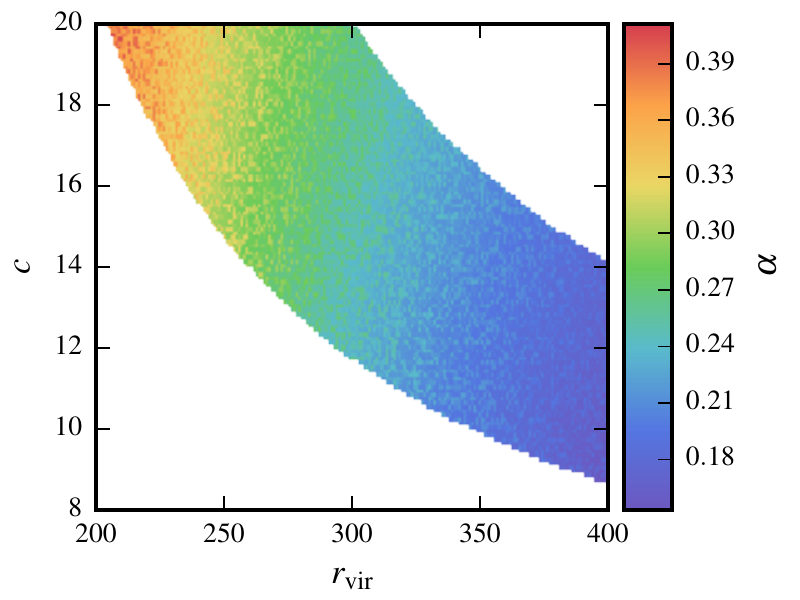}
    \includegraphics[width=\linewidth]{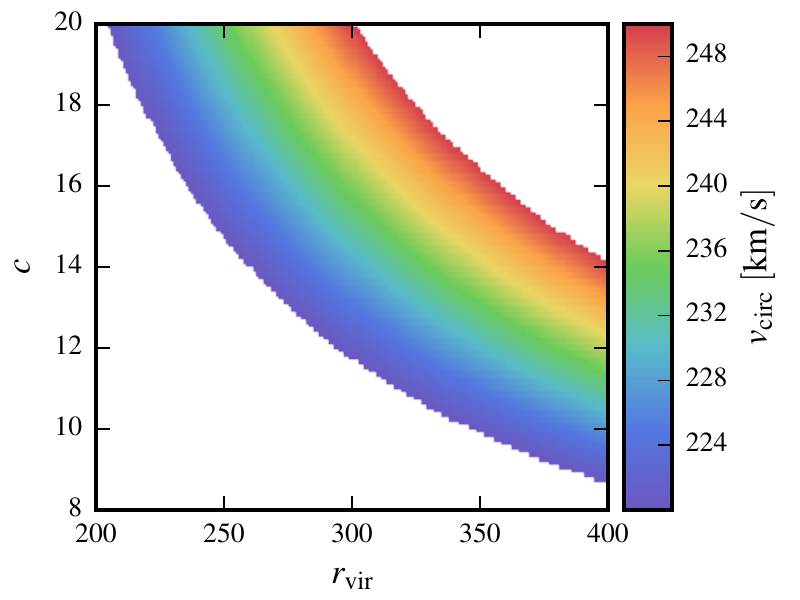}
    \caption{The variation in potential slope (upper panel) and circular velocity at the solar radius (lower panel) for a grid of halo models assumed to be NFW in shape and defined by a concentration $c$ and a virial radius $r_\mathrm{vir}$. We only show halos with circular velocities at the solar radius $R_\odot = 8.29$~kpc of $220 \le \vcirc \le 250$~\kms\ to force consistency with observations. The white regions are halos with circular velocities outside of this range.}
    \label{figure:halogrid}
\end{figure}

\begin{figure}
    \centering
    \includegraphics[width=0.8\linewidth]{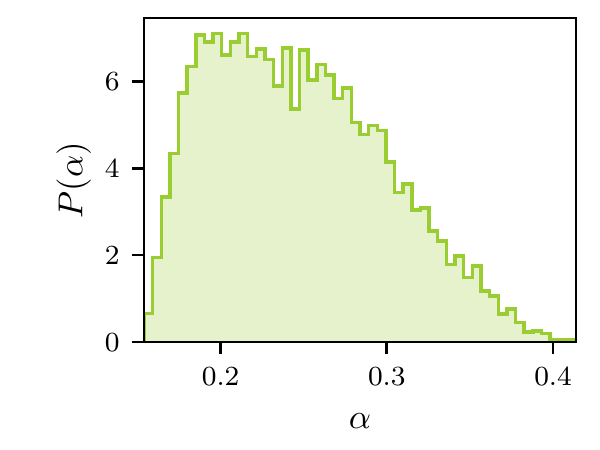}
    \caption{Distribution of $\alpha$ values fitted to the grid of sample halos; halos for which the circular velocity at the solar radius is inconsistent with observations have been removed.}
    \label{figure:alphas}
\end{figure}

\autoref{figure:halogrid} shows the variation in $\alpha$ (upper panel) and $\vcirc$ (lower panel) across our halo sample, as indicated by the respective colour bars. The white regions are halos with circular velocities inconsistent with observations. The corresponding distribution of $\alpha$ values is shown in \autoref{figure:alphas}.

\subsection{Monte Carlo Simulations}
\label{section:mcsims}

The variation in the mass estimates for different halos assesses our uncertainty in the particular density profile of the halo, but does not assess how well the mass estimators themselves are able to recover the true mass of the MW using \Nsample\ tracers drawn from the underlying distribution. We do this using the set of Monte Carlo simulations described in \citet{Watkins2010}. Briefly, the simulations create a set of tracer objects drawn from a power-law density with index $\gamma$, with velocities consistent with a power-law potential with index $\alpha$ and with anisotropy $\beta$. We then use the TME to estimate the mass $\Mtme$ within $\rmax$, and compare this estimated mass with the known true mass of the simulation $\Mtrue$ within $\rmax$.

For these simulations, we use density slope $\gamma = \valgamma$ (as calculated in \autoref{section:density}), and anisotropy $\beta = \valbeta$ (the median value found in \autoref{section:anisotropy}). For the potential slope $\alpha$, we select a `typical' halo using the halo grid described in \autoref{section:potential}: we choose to use $\rvir = 300$~kpc (the middle value of the range sampled), and then use the grid to find the value of concentration $c$ for which the circular velocity at the solar radius is closest to the observed value of $V_\odot$, which results in $c = \valconc$. We use the value of $\alpha = \valalpha$ calculated for this halo. We further use the observed solar radius and the circular velocity at the solar radius for the fiducial radius and fiducial velocity in the potential power-law. Although our analysis does use a range of values for both $\alpha$ and $\beta$, using fixed values for our simulations is sufficient for our purposes here, which is to assess how far our mass estimate based on a single cluster sample may be from the true value, as we do not expect the performance of the TME with \Nsample\ tracers to change with $\alpha$ or $\beta$.

\begin{figure}
    \centering
    \includegraphics[width=0.8\linewidth]{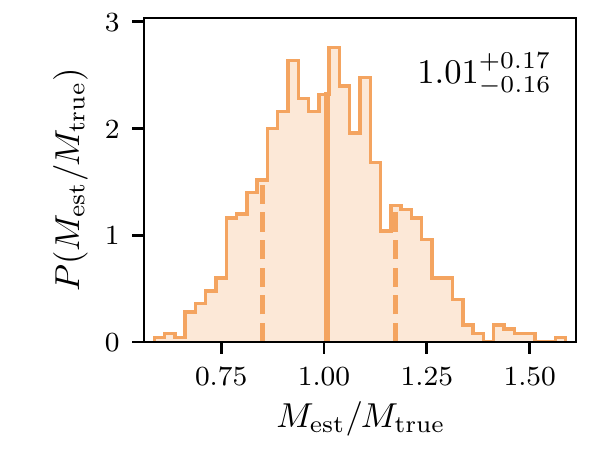}
    \caption{Performance of the TME using \Nsample\ tracers. We ran 1000 Monte Carlo simulations (see text for details). Here we show a histogram of the ratio $\Mtme/\Mtrue$ of the estimated mass to the true mass for the simulations, with the median value marked with a solid line and the 15.9 and 84.1 percentiles shown as dashed lines. These values are all given in the upper-right corner. We see that on average the estimator does indeed recover the true mass. We incorporate the scatter in the recovered values into the uncertainties for our final mass estimate.}
    \label{figure:mc_tme_mass}
\end{figure}

We generate 1000 simulations of \Nsample\ clusters, from which we find that the ratio of the estimated mass to the true mass $f = \Mtme/\Mtrue = \valf^{+\valepf}_{-\valemf}$, suggesting that the estimators are able to recover the true mass on average, assuming our models are a good description of nature. The full distribution of $f$ values is shown in \autoref{figure:mc_tme_mass} with the median and 15.9 and 84.1 percentiles marked as solid and dashed lines. We will later use the results of these simulations to ensure that the scatter in the recovery of the true mass from a single sample is accurately accounted for in our uncertainties.

\subsection{Tracer Mass Estimates}
\label{section:tmes}

Now that we have estimated the anisotropy $\beta$, density power-law index $\gamma$, and potential power-law indices $\alpha$, we can combine this information with the cluster position and velocity data, and finally estimate the mass of the Milky Way within $\rmax = \valrmax$~kpc using \autoref{eqn:TME}. From this mass, we can also estimate the circular velocity of the MW at $\rmax$ as
\begin{equation}
    \vrmax = \sqrt{\frac{G \Mrmax}{\rmax}} .
\end{equation}

We begin with the set of halos generated in \autoref{section:potential} with circular velocities at the solar radius consistent with observations. We previously estimated $\alpha$ values for every halo. Additionally, we draw an anisotropy $\beta$ at random from the posterior distribution of anisotropy values calculated in \autoref{section:anisotropy} for each halo. The value of $\gamma$ is fixed from \autoref{section:density} and is the same for every halo. Finally, we draw positions and velocities from the distributions for each cluster and use \autoref{eqn:TME} to estimate the mass $\Mtme$ for each halo in the grid.

We know from the Monte Carlo simulations in \autoref{section:mcsims}, that there is some scatter in the performance of the estimator for a single sample of \Nsample\ clusters and that the true mass is related to the estimated mass via $\Mtrue = \Mtme/f$. For each sample halo, we draw a value of $f$ at random from the posterior distribution calculated in \autoref{section:mcsims} and use that to infer the true mass from the TME mass. We adopt the median of the resulting distribution as the best mass estimate, and use the 15.9 and 84.1 percentiles to estimate uncertainties. Thus, we estimate the mass of the MW within $\valrmax$~kpc to be
\begin{equation}
    M(<\valrmax \; \mathrm{kpc}) = \valMrmax^{+\valepMrmax}_{-\valemMrmax} \times 10^{12} \mathrm{M_\odot}
    \label{eq:earlieres}
\end{equation}
and the circular velocity of the MW at this distance to be
\begin{equation}
    \vcirc (\valrmax \; \mathrm{kpc}) = \valvrmax^{+\valepvrmax}_{-\valemvrmax} \mathrm{km\,s^{-1}} .
\end{equation}
Note that the scatter in the performance of the estimator for a single sample and our uncertainties on both $\alpha$, $\beta$, and the cluster properties have been naturally folded into our results using this method.

The TME only allows us to estimate the mass inside the radius of the outermost cluster in our sample. However, we can use our halo grid to predict what the virial mass\footnote{We calculate the virial mass at each grid point directly given the NFW parameters of the point. The virial mass is defined as the mass inside the virial radius, and the virial radius is defined as the radius at which the mean overdensity of the halo relative to the critical density is $\Delta_\mathrm{vir}$. We adopt the prescription for $\Delta_\mathrm{vir}$ from \citet{Bryan1998}. Note that we do not estimate the virial mass from the power-law fits, as the power laws are assumed to hold over the range of the GC data, not out into the outskirts of the halo.} of the MW might be, given the value of $\Mrmax$ we have estimated.

\begin{figure}
    \centering
    \includegraphics[width=\linewidth]{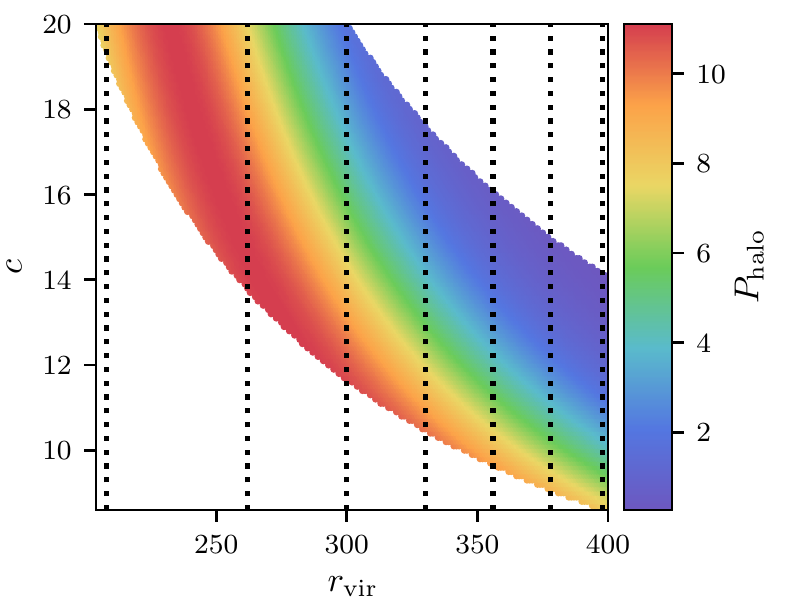}
    \caption{The probability $P_\mathrm{halo}$ of a set of NFW models defined by concentration $c$ and virial radius $r_\mathrm{vir}$, based on the $\Mgridrmax$ predicted for the model and the actual $\Mrmax$ measured from the data. White regions are halos previously rejected as they have circular velocities inconsistent with observed values. The vertical lines mark, from left to right, halos with virial masses of $(0.5,1.0,1.5,2.0,2.5,3.0,3.5) \times 10^{12}$~\Msun.}
    \label{figure:Phalo}
\end{figure}

For each of the sample halos in the grid, we can estimate the mass $\Mgridrmax$ of the MW inside $\rmax$ for the halo model. We then define the probability density $P_\mathrm{halo}$ of the model value, given the distribution of $\Mrmax$ we estimate from the data. The variation of $P_\mathrm{halo}$ over the halo grid is shown in \autoref{figure:Phalo}. Halos on the left side of the swath in the diagram are most consistent with the observations. For each halo in the grid, we can also estimate the mass inside the virial radius $\Mgridvir$. Now we ask what are the virial masses of the halos for which $\Mgridrmax$ agrees with our measured value $\Mrmax$? That is, we look at the distribution of virial masses over the grid, weighted by the $P_\mathrm{halo}$ values that assesses the consistency of that grid point with our measured value. From this, we can calculate the expected value and 15.9 and 84.1 percentiles of the virial mass implied by our tracer mass estimates, taking these probabilities into account. Thus, we find the virial mass of the Milky Way to be
\begin{equation}
    \Mvir = \valMvir^{+\valepMvir}_{-\valemMvir} \times 10^{12} \mathrm{M_\odot} .
\end{equation}
Again, the scatter inherent in the estimator and the uncertainties on the $\alpha$ and $\beta$ have been propagated into this result. Our estimate for $\Mvir$ has a larger fractional uncertainty than $\Mrmax$ owing to the uncertainty in the extrapolation of the dark halo mass profile to large radii.

\begin{figure*}
    \centering
    \includegraphics[width=0.32\linewidth]{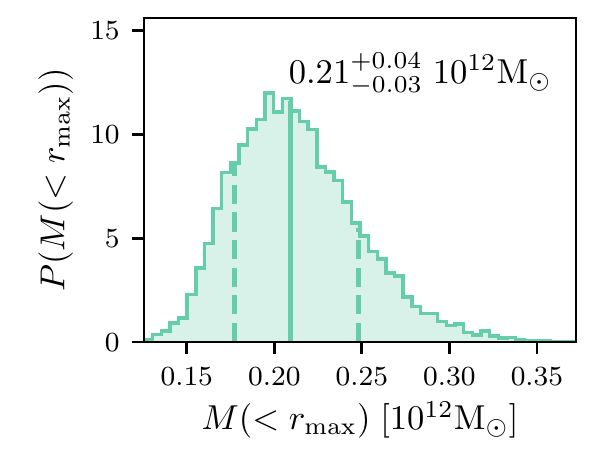}
    \includegraphics[width=0.32\linewidth]{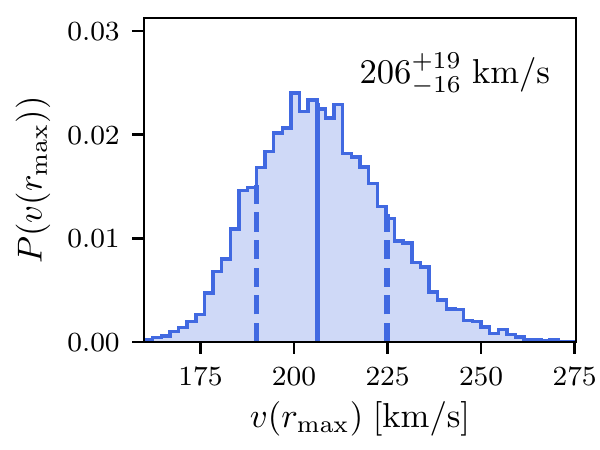}
    \includegraphics[width=0.32\linewidth]{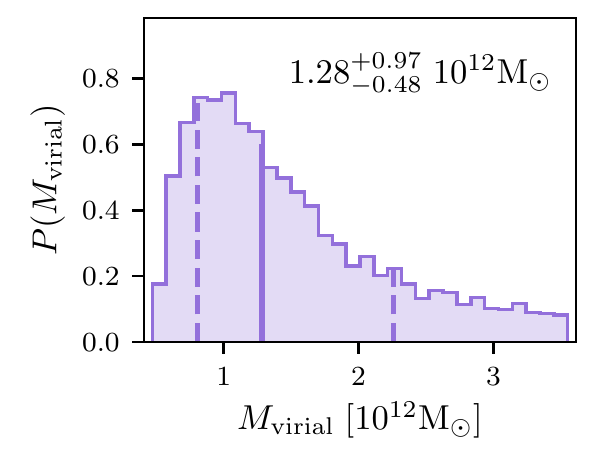}
    \caption{Summary of results using the Gaia GC sample for which $\rmax = \valrmax$~kpc. Left: Distribution of tracer mass estimates $\Mrmax$. Middle: Distribution of estimates of the circular velocity $\vrmax$ made using the mass estimates in the left panel. Right: Virial mass $\Mvir$ estimates inferred from the grid of halos sampled, weighted by the match to distribution of masses in the left panel. In each case, we adopt the median of the distribution and the 15.9 and 84.1 percentiles as the estimate and its uncertainties; these are given in the top-right corner of each panel.}
    \label{figure:halo_histograms}
\end{figure*}
\autoref{figure:halo_histograms} summarises our results. The left panel shows the distribution of mass estimates $M(<\valrmax \; \mathrm{kpc})$; the middle panel shows the distribution of $\vcirc (\valrmax \; \mathrm{kpc})$ estimates; and the right panel shows the resulting distribution of predicted virial masses $\Mvir$.

\subsection{Expanded GC Sample}
\label{section:expanded}

Our analysis so far has been limited to $\valrmax$~kpc, and only probes the very inner parts of the halo. Furthermore, the uncertainties on our virial mass are large due to extrapolation of our results out to significantly larger radii. More clusters further out in the halo would thus be extremely beneficial. Increased sample size would also help to decrease uncertainties in all estimates, assuming our model assumptions are reasonable, or highlight problems if our model assumptions are incorrect.

As described, in \autoref{section:HSTclusters}, we augment our sample with 12 extra clusters with \textit{HST} PMs from \citet{Sohn2018}. We repeat the analysis as laid out above. The density slope $\gamma$ remains the same, but all other parts of the analysis necessarily change with the new sample. We find an anisotropy,
\begin{equation}
	\beta = \valbetacomb^{+\valepbetacomb}_{-\valembetacomb} .
\end{equation}
The Monte Carlo simulations are run with the same $\alpha$ and $\gamma$ as before, but with the revised $\beta$ and with an increased radial range and give $f = \valfcomb^{+\valepfcomb}_{-\valemfcomb}$. For the new $\rmax = \valrmaxcomb$~kpc, we then estimate
\begin{equation}
	M(<\valrmaxcomb \; \mathrm{kpc}) = \valMrmaxcomb^{+\valepMrmaxcomb}_{-\valemMrmaxcomb}  \times 10^{12}\,\mathrm{M_\odot} ,
\end{equation}
which corresponds to a circular velocity at $\rmax$ of
\begin{equation}
	\vcirc (\valrmaxcomb \; \mathrm{kpc}) = \valvrmaxcomb^{+\valepvrmaxcomb}_{-\valemvrmaxcomb} \mathrm{km\,s^{-1}} .
\end{equation}
We again extrapolate out to estimate the virial mass and find
\begin{equation}
	\Mvir = \valMvircomb^{+\valepMvircomb}_{-\valemMvircomb}  \times 10^{12}\,\mathrm{M_\odot} .
\end{equation}

\begin{deluxetable*}{ccccccc}
\tablecaption{Summary of mass results for the halo GC subsamples. \label{table:results_masses}}
\tablehead{
   \colhead{Sample} & 
   \colhead{$N$} & 
   \colhead{$\rmin$} & 
   \colhead{$\rmax$} &
   \colhead{$\Mrmax$} & 
   \colhead{$\vrmax$} &
   \colhead{$\Mvir$} \\
   \colhead{} & 
   \colhead{} & 
   \colhead{(kpc)} & 
   \colhead{(kpc)} & 
   \colhead{$10^{12}$~\Msun} & 
   \colhead{(\kms)} &
   \colhead{$10^{12}$~\Msun}
}
\startdata
A\tablenotemark{a} & 34 & 2.0 & 21.1 & $0.21^{+0.04}_{-0.03}$ & $206^{+19}_{-16}$ & $1.28^{+0.97}_{-0.48}$ \\
B\tablenotemark{b} & 46 & 2.0 & 39.5 & $0.42^{+0.07}_{-0.06}$ & $214^{+17}_{-15}$ & $1.54^{+0.75}_{-0.44}$ \\
C\tablenotemark{c} & 26 & 2.0 & 21.1 & $0.21^{+0.04}_{-0.03}$ & $209^{+21}_{-17}$ & $1.34^{+1.02}_{-0.50}$ \\
D\tablenotemark{d} & 41 & 4.1 & 39.5 & $0.36^{+0.07}_{-0.06}$ & $199^{+19}_{-16}$ & $1.22^{+0.63}_{-0.36}$ \\
\enddata
\tablenotetext{a}{\textit{Gaia} GCs.}
\tablenotetext{b}{\textit{Gaia} and \textit{HST} GCs.}
\tablenotetext{c}{\textit{Gaia} GCs with possible merger group removed.}
\tablenotetext{d}{\textit{Gaia} and \textit{HST} GCs with high $\vtan$ clusters removed.}
\end{deluxetable*}

We summarise our results in \autoref{table:results_masses}. Sample A is the set of \textit{Gaia} measurements and Sample B is the set of combined \textit{Gaia} and \textit{HST} measurements. Samples C and D are discussed in \autoref{section:discussion}.

\section{Discussion}
\label{section:discussion}

\subsection{Anisotropy}
\label{section:discussaniso}

Using only the \textit{Gaia} halo GCs we estimated anisotropy $\beta = \valbeta^{+\valepbeta}_{-\valembeta}$ over \rrange, and using the expanded sample we found $\beta = \valbetacomb^{+\valepbetacomb}_{-\valembetacomb}$ over \rrangecomb. Both values indicate that the halo over the range of the GC sample is radially anisotropic. \citet{Sohn2018} reported $\beta = 0.609^{+0.130}_{-0.229}$ over the range $10.6 < r < 39.5$~kpc. All of these values are consistent within their uncertainties, but the trends suggest that the halo becomes more radially anisotropic in its outer regions, in line with predictions from cosmological simulations \citep[e.g.][]{diemand2007}.

There are a number of other estimates for the anisotropy of the halo over radial ranges that overlap that of our sample. Our radial $\beta$ is inconsistent with the estimates for individual halo star samples of \citet{Sirko2004} and \citet{Cunningham2016}, both of which favour an isotopic or even tangentially-anisotropic halo (although the uncertainties in the latter were large enough that that results cannot be called truly discrepant), but in good agreement with the radial estimates from \citet{Bond2010} and \citet{Deason2012}.

\subsection{Masses}
\label{section:discussmass}

As we discussed in \autoref{section:introduction}, MW mass estimates can vary markedly based on the types of data used, the techniques used, and the assumptions that go into the mass estimate, with estimates for the total mass of the MW varying between $\sim 0.5 - 3 \times 10^{12}$\Msun. Furthermore, only the timing argument and abundance-matching studies actually estimate the total mass of the MW, most other estimates can only measure the mass within the extents of the dataset being used, as we have done here, so most estimates are given at different radii, which makes them hard to compare.

Indeed, our mass estimates at $\valrmax$~kpc and $\valrmaxcomb$~kpc cannot easily be compared directly. We do note that the latter mass is about twice as large for about twice the size of the enclosed radius, consistent with naive expectations for a nearly isothermal sphere. Our values are instead most usefully compared to other estimates near these radii.

\citet{Kafle2012} estimated $M(<25 \; \mathrm{kpc}) \approx 0.21 \times 10^{12}$~\Msun, unfortunately without uncertainties, by analysing blue horizontal branch stars in the halo, and \citet{Kuepper2015} estimated $M(<19 \; \mathrm{kpc}) = 0.21 \pm 0.04 \times 10^{12}$~\Msun\ by analysing the orbit of the Palomar 5 tidal stream, both in good agreement with our estimate of $M(<\valrmax \; \mathrm{kpc}) = \valMrmax^{+\valepMrmax}_{-\valemMrmax} \times 10^{12} \mathrm{M_\odot}$. Since completing this work, there have been further mass estimates from Gaia measurements: \citet{Posti2019} applied a Bayesian estimator to the Gaia sample as we used here to estimate $M(<20 \; \mathrm{kpc}) = 0.191^{+0.017}_{-0.015} \times 10^{12}$~\Msun; and \citet{Eadie2018} applied an alternative Bayesian estimator to the expanded sample from 
\citet{Vasiliev2019} to estimate $M(<25 \; \mathrm{kpc}) = 0.26^{+0.03}_{-0.02} \times 10^{12}$~\Msun, also in good agreement with our results.

\citet{Sohn2018} estimated $M(<39.5 \; \mathrm{kpc}) = 0.60^{+0.17}_{-0.11} \times 10^{12}$~\Msun\ using the same a set of \textit{HST} clusters we used to augment our sample, slightly higher but still consistent with our estimate here of $M(<\valrmaxcomb \; \mathrm{kpc}) = \valMrmaxcomb^{+\valepMrmaxcomb}_{-\valemMrmaxcomb} \times 10^{12} \mathrm{M_\odot}$.

There are a number of previous estimates of the mass within 50~kpc including $0.54^{+0.02}_{-0.36} \times 10^{12}$~\Msun\ \citep{Wilkinson1999}, $0.55^{+0.00}_{-0.02} \times 10^{12}$~\Msun\ \citep{Sakamoto2003}, and $0.42 \pm 0.04 \times 10^{12}$~\Msun\ \citep{Deason2012}. Remembering that these estimates were made $\sim$10~kpc further out than our sample, the two former estimates are in extremely good agreement with our estimate given expectations for a nearly isothermal sphere, and the latter is slightly lower, but still consistent with our estimate.

More recently, \citet{Vasiliev2019} estimated $M(<50 \; \mathrm{kpc}) = 0.6^{+0.14}_{-0.09} \times 10^{12}$~\Msun\ and \citet{Eadie2018} estimated $M(<50 \; \mathrm{kpc}) = 0.37^{+0.04}_{-0.03} \times 10^{12}$~\Msun, both using an expanded sample of PMs measured with Gaia from the former work. Our estimate is consistent with the \citet{Vasiliev2019} result and marginally inconsistent with the \citet{Eadie2018} result; our estimate further in at $\sim$21~kpc was consistent with the \citet{Eadie2018} estimate at 25~kpc, suggesting that their method predicts a steeper mass density slope than favoured by our models.

We also note that \citet{Gibbons2014} estimated a mass $0.41 \pm 0.04 \times 10^{12}$~\Msun\ inside 100~kpc by fitting to the Sagittarius tidal stream, clearly at odds with our results. These authors only fitted the locations of the apocentres and pericentres of the trailing and leading arms of the Sagittarius, and not the intervening locations. The dataset is now much richer, with \textit{Gaia} DR2 likely to add to our knowledge of the proper motions along the stream. It would be interesting to re-visit the work of \citet{Gibbons2014} in the light of this.

Now let us consider our virial mass estimates. Using only the \textit{Gaia} GCs, we find $\valMvir^{+\valepMvir}_{-\valemMvir} \times 10^{12}\,\mathrm{M_\odot}$, and using the expanded sample, we find $\valMvircomb^{+\valepMvircomb}_{-\valemMvircomb} \times 10^{12}\,\mathrm{M_\odot}$. Both of these values are in good agreement, but note that the error bars on the latter are smaller than the error bars on the former. The increased sample size may have a small effect here, but the main contribution to this decrease comes from the fact that there is less extrapolation involved in estimating a virial mass from data at $\sim$40~kpc than there is from data at $\sim$20~kpc. This means that there is less variation in the allowed halos and in turn allows us to constrain the virial mass much better.

\citet{Sohn2018} estimated a virial mass of $1.87^{+0.67}_{-0.47} \times 10^{12}$~\Msun, slightly higher than but in good agreement with both of our values.

Comparison with other studies is again tricky as the definition of 'virial' can change from study to study. The recent review by \citet{BlandHawthorn2016} put a number of different estimates at a different radii onto the same scale for comparison. Our estimates are best compared against their $M_{100}$ values, that is the mass within the radius for which the mean overdensity is 100, which is very close to the mean overdensity value we use in this work. Our estimates are inconsistent with the most massive and least massive of these, and agree best with the intermediate values. (See also Figure 1 of \citet{Wang2015} and Figure 1 of \citet{Eadie2017} for further comparisons.)

More recently, \citet{Vasiliev2019} estimated a virial mass $0.8^{+0.5}_{-0.2} \times 10^{12}$~\Msun using a large sample of GCs with Gaia PMs. This is considerably smaller than our value, and the virial radius they estimate ($\sim 160 \pm 20$~kpc) is much smaller than those our method favours, implying that the outer halo density is steeper than the NFW value of $-3$. Many of the additional clusters in the \citet{Vasiliev2019} dataset are further out in the halo than the sample we used, which in principle should give stronger constraints on the mass at larger distances. However, these outer clusters are more likely to have been recently accreted and may not be fully phase-mixed, which could invalidate any mass estimates. To fully explore these differences and their implications is beyond the scope of this paper.

\subsection{Halo Density Profile}
\label{section:discusshalos}

We can also consider the implications of our results for the density profile of the halo. \autoref{figure:Phalo} shows the agreement between the mass enclosed at $\rmax$ for the model and from the data for the set of allowed halos with consistent circular velocities at the solar radius. On this, we mark the virial radii for which the halo virial masses are, from left to right, $0.5-3.5 \times 10^{12}$~\Msun at $0.5 \times 10^{12}$~\Msun intervals for reference. It is clear that the data can be explained by either low concentration, high virial radius (and thus high virial mass) halos -- similar to those favoured by cosmological simulations \citep{Klypin2011}\footnote{Note though that these simulations do not fully consider the effect of baryons in shaping the halo.}
 -- or by high concentration, low virial radius/mass halos. The latter seem to be favoured over the former, as there are more such halos that can adequately fit the data, but we cannot rule out either.

To derive our results, we have made a number of assumptions, one of which is that the halo and the GC distributions are spherical, neither of which is necessarily true. Indeed, there are some hints of non-sphericity in that $\sigma_\theta < \sigma_\phi$ (see \autoref{table:results_kinematics}). To lowest order, the predictions of a spherical model, when applied to non-spherical distributions, can be interpreted as an estimate of the spherically averaged quantities of the actual distribution. In reality, some bias may be introduced. The exact size of this needs to be quantified through analysis of either, e.g., triaxial equilibrium models, or cosmological simulations with non-spherical distributions, both of which are outside the scope of the present paper.

\subsection{Substructure}

One of the assumptions in our work is that the halo GCs comprise a statistically independent, well-mixed tracer population in the MW's gravitational potential. It is for this reason that we excluded all-but-one of the GCs associated with the Sagittarius dSph. However, there have been recent claims that up to two thirds of the local stellar halo may have been deposited by the single encounter of a massive dwarf galaxy~\citep[e.g.,][]{Deason2013, Belokurov2018, Myeong2018}. The evidence for this rests mainly on the highly radially anisotropic, non-Gaussian distribution of velocities of metal-rich halo stars in both the SDSS-\textit{Gaia} and \textit{Gaia} DR2 catalogues. If true, then such a massive satellite will also have been accompanied by its own retinue of GCs. Very recently, \citet{Myeong2018b} have tentatively used \textit{Gaia} DR2 to identify 8 halo GCs associated with this merger event (NGCs 1851, 1904, 2298, 2808, 5286, 6779, 6864, and 7089). 

It is prudent to check the robustness of our results to these claims. Re-running the calculations for our DR2-only sample with these 8 GCs removed, we obtain
\begin{equation}
    M(<\valrmax \; \mathrm{kpc}) = 0.21^{+0.04}_{-0.03} \times 10^{12} \mathrm{M_\odot},
\end{equation}
very close to our earlier result in eq.~(\ref{eq:earlieres}). We note that while the mass enclosed hardly changes, the inferred anisotropy does
\begin{equation}
	\beta = 0.24^{+0.23}_{-0.31} .
\end{equation}
This change is understandable, as the most eccentric GCs have been excised from the sample. Indeed, \citet{Myeong2018b} identified their 8 GCs from clustering in radial action, and so the removed GCs do make a prominent contribution to the anisotropy parameter. These results are added to \autoref{table:results_kinematics} and \autoref{table:results_masses} as Sample C.

Alternatively, recently accreted GCs or young halo GCs that have not yet phase-mixed with the rest of the GC population could be on highly tangential orbits. As we noted in \autoref{section:anisotropy}, we see that the \textit{Gaia} (and \textit{HST}) GCs show net tangential motion; this is also apparent in \autoref{figure:data} where the high $\vtan$ GCs have been coloured in orange and cyan. To assess the dependence of our results on these outliers, we reran the analysis of the combined \textit{Gaia}+\textit{HST} sample with these clusters removed. We now obtain an anisotropy estimate of
\begin{equation}
	\beta = 0.76^{+0.06}_{-0.08} .
\end{equation}
The value of $\beta$ is now obviously more radially anisotropic than before as we have removed high $\vtan$ clusters. The enclosed mass within $\rmax$ is becomes
\begin{equation}
    M(<\valrmaxcomb \; \mathrm{kpc}) = 0.36^{+0.07}_{-0.06} \times 10^{12} \mathrm{M_\odot},
\end{equation}
and the inferred virial mass is
\begin{equation}
    \Mvir = 1.18^{+0.51}_{-0.34} \times 10^{12} \mathrm{M_\odot} .
\end{equation}
These results are added to \autoref{table:results_kinematics} and \autoref{table:results_masses} as Sample D. The inferred masses are somewhat decreased, but still, within the uncertainty ranges of the previously quoted values. These tests with modified samples add confidence that our inferred MW mass estimates are pleasingly robust to potential substructure in the halo GC distribution.

\section{Conclusions}
\label{section:conclusions}

We have used Galactocentric motions for a set of $\Nsample$ halo GCs estimated using PM data from the second Gaia data release, to estimate the anisotropy of the halo GC population and then to estimate the mass of the MW inside $\valrmax$~kpc, the position of the most distant cluster in our sample. Combined with a catalogue of clusters from a recent \textit{HST} study, we were able to estimate the anisotropy and mass of the MW inside $\valrmaxcomb$~kpc.

Using only the \textit{Gaia} sample, we find an anisotropy $\beta = \valbeta^{+\valepbeta}_{-\valembeta}$ in the range of the clusters \rrange. With the addition of the \textit{HST} clusters, we find $\beta = \valbetacomb^{+\valepbetacomb}_{-\valembetacomb}$ over \rrangecomb. This suggests that the halo is radially anisotropic, consistent with a number of previous estimates and predictions from cosmological simulations and in good agreement with a number of similar studies, including the study of 16 halo GCs with \textit{HST} by \cite{Sohn2018}.

We estimate the masses $M(<\valrmax \; \mathrm{kpc}) = \valMrmax^{+\valepMrmax}_{-\valemMrmax} \times 10^{12} \mathrm{M_\odot}$ and $M(<\valrmaxcomb \; \mathrm{kpc}) = \valMrmaxcomb^{+\valepMrmaxcomb}_{-\valemMrmaxcomb} \times 10^{12} \mathrm{M_\odot}$. These masses correspond to circular velocities of $\vcirc (\valrmax \; \mathrm{kpc}) = \valvrmax^{+\valepvrmax}_{-\valemvrmax} \mathrm{km\,s^{-1}}$ and $\vcirc (\valrmaxcomb \; \mathrm{kpc}) = \valvrmaxcomb^{+\valepvrmaxcomb}_{-\valemvrmaxcomb} \mathrm{km\,s^{-1}}$ respectively, which, compared with estimates for the circular velocity at the solar radius and to each other, is consistent with a rotation curve that is not falling rapidly over the radial range of our data.

From these, we are also able to place constraints on the virial mass of the MW. We favour the results for the combined sample here, as it is more reasonable to perform this extrapolation for a larger sample size with a broader radial range and, more importantly, greater reach. We find $\Mvir = \valMvircomb^{+\valepMvircomb}_{-\valemMvircomb} \times 10^{12} \mathrm{M_\odot}$, again of intermediate size compared with previous estimates. All of our mass estimates are intermediate in value when compared to the range of values found in the literature, with both low-mass ($< 10^{12}$~\Msun) and very high mass ($\gtrsim 2.5 \times 10^{12}$~\Msun) MWs generally disfavoured.

Previous mass estimates have often been limited by either the mass-anisotropy degeneracy for LOS velocity samples, or the small samples sizes for distant objects with 3D motions. Given the new results from \textit{Gaia} and \textit{HST}, these are now both resolved. Various kinds of systematics may now become the dominant source of uncertainty. Nevertheless, further progress will come from having yet larger sample sizes. \citet{GaiaGCs} measured PMs for only 75 GCs out of the 157 known in the MW \citep[][2010 edition]{Harris1996} by making extremely conservative cuts on the number of member stars identified. It is likely that PMs and, hence, Galactocentric motions can be measured for many more Galactic GCs, both using DR2 and future data releases. Such measurements will further refine our understanding of the MW mass.

\acknowledgments

LLW wishes to thank Alis Deason, Mark Fardal, Elena Pancino, and Mark Gieles for very useful conversations related to this work. We also thank Amina Helmi for making supporting data for \citet{GaiaGCs} available electronically.\footnote{\url{https://www.astro.rug.nl/~ahelmi/research/dr2-dggc/}} We thank the referee for the suggestions that improved the presentation of our results. Support for this work was provided by grants for \textit{HST} program GO-14235 provided by the Space Telescope Science Institute, which is operated by AURA, Inc., under NASA contract NAS 5-26555.

This work has made use of data from the European Space Agency (ESA) mission \textit{Gaia}\footnote{\url{https://www.cosmos.esa.int/gaia}}, processed by the \textit{Gaia} Data Processing and Analysis Consortium (DPAC).\footnote{\url{https://www.cosmos.esa.int/web/gaia/dpac/consortium}} Funding for the DPAC has been provided by national institutions, in particular the institutions participating in the \textit{Gaia} Multilateral Agreement.

This research made use of Astropy\footnote{\url{http://www.astropy.org}}, a community-developed core Python package for Astronomy. This research has made use of NASA's Astrophysics Data System Bibliographic Services.

This project is part of the HSTPROMO (High-resolution Space Telescope PROper MOtion) Collaboration\footnote{\url{http://www.stsci.edu/~marel/hstpromo.html}}, a set of projects aimed at improving our dynamical understanding of stars, clusters and galaxies in the nearby Universe through measurement and interpretation of proper motions from \textit{HST}, \textit{Gaia}, and other space observatories. We thank the collaboration members for the sharing of their ideas and software.


\vspace{5mm}
\facility{Gaia, HST}


\software{
	astropy \citep{astropy2013, astropy2018},
	\textsc{emcee} \citep{ForemanMackey2013}
}

\bibliography{refs}

\begin{thebibliography}{}
\expandafter\ifx\csname natexlab\endcsname\relax\def\natexlab#1{#1}\fi
\providecommand{\url}[1]{\href{#1}{#1}}

\bibitem[{{Anderson} \& {King}(2003)}]{Anderson2003}
{Anderson}, J., \& {King}, I.~R. 2003, \aj, 126, 772

\bibitem[{{Annibali} {et~al.}(2018){Annibali}, {Morandi}, {Watkins}, {Tosi},
  {Aloisi}, {Buzzoni}, {Cusano}, {Fumana}, {Marchetti}, {Mignoli},
  {Mucciarelli}, {Romano}, \& {van der Marel}}]{Annibali2018}
{Annibali}, F., {Morandi}, E., {Watkins}, L.~L., {et~al.} 2018, \mnras, 476,
  1942

\bibitem[{{Astropy Collaboration} {et~al.}(2013){Astropy Collaboration},
  {Robitaille}, {Tollerud}, {Greenfield}, {Droettboom}, {Bray}, {Aldcroft},
  {Davis}, {Ginsburg}, {Price-Whelan}, {Kerzendorf}, {Conley}, {Crighton},
  {Barbary}, {Muna}, {Ferguson}, {Grollier}, {Parikh}, {Nair}, {Unther},
  {Deil}, {Woillez}, {Conseil}, {Kramer}, {Turner}, {Singer}, {Fox}, {Weaver},
  {Zabalza}, {Edwards}, {Azalee Bostroem}, {Burke}, {Casey}, {Crawford},
  {Dencheva}, {Ely}, {Jenness}, {Labrie}, {Lim}, {Pierfederici}, {Pontzen},
  {Ptak}, {Refsdal}, {Servillat}, \& {Streicher}}]{astropy2013}
{Astropy Collaboration}, {Robitaille}, T.~P., {Tollerud}, E.~J., {et~al.} 2013,
  \aap, 558, doi:10.1051/0004-6361/201322068

\bibitem[{{Astropy Collaboration} {et~al.}(2018){Astropy Collaboration},
  {Price-Whelan}, {Sip{\'{o}}cz}, {G{\"u}nther}, {Lim}, {Crawford}, {Conseil},
  {Shupe}, {Craig}, {Dencheva}, {Ginsburg}, {VanderPlas}, {Bradley},
  {P{\'e}rez-Su{\'a}rez}, {de Val- Borro}, {Aldcroft}, {Cruz}, {Robitaille},
  {Tollerud}, {Ardelean}, {Babej}, {Bach}, {Bachetti}, {Bakanov}, {Bamford},
  {Barentsen}, {Barmby}, {Baumbach}, {Berry}, {Biscani}, {Boquien}, {Bostroem},
  {Bouma}, {Brammer}, {Bray}, {Breytenbach}, {Buddelmeijer}, {Burke},
  {Calderone}, {Cano Rodr{\'\i}guez}, {Cara}, {Cardoso}, {Cheedella}, {Copin},
  {Corrales}, {Crichton}, {D'Avella}, {Deil}, {Depagne}, {Dietrich}, {Donath},
  {Droettboom}, {Earl}, {Erben}, {Fabbro}, {Ferreira}, {Finethy}, {Fox},
  {Garrison}, {Gibbons}, {Goldstein}, {Gommers}, {Greco}, {Greenfield},
  {Groener}, {Grollier}, {Hagen}, {Hirst}, {Homeier}, {Horton}, {Hosseinzadeh},
  {Hu}, {Hunkeler}, {Ivezi{\'c}}, {Jain}, {Jenness}, {Kanarek}, {Kendrew},
  {Kern}, {Kerzendorf}, {Khvalko}, {King}, {Kirkby}, {Kulkarni}, {Kumar},
  {Lee}, {Lenz}, {Littlefair}, {Ma}, {Macleod}, {Mastropietro}, {McCully},
  {Montagnac}, {Morris}, {Mueller}, {Mumford}, {Muna}, {Murphy}, {Nelson},
  {Nguyen}, {Ninan}, {N{\"o}the}, {Ogaz}, {Oh}, {Parejko}, {Parley}, {Pascual},
  {Patil}, {Patil}, {Plunkett}, {Prochaska}, {Rastogi}, {Reddy Janga},
  {Sabater}, {Sakurikar}, {Seifert}, {Sherbert}, {Sherwood-Taylor}, {Shih},
  {Sick}, {Silbiger}, {Singanamalla}, {Singer}, {Sladen}, {Sooley},
  {Sornarajah}, {Streicher}, {Teuben}, {Thomas}, {Tremblay}, {Turner},
  {Terr{\'o}n}, {van Kerkwijk}, {de la Vega}, {Watkins}, {Weaver}, {Whitmore},
  {Woillez}, {Zabalza}, \& {Astropy Contributors}}]{astropy2018}
{Astropy Collaboration}, {Price-Whelan}, A.~M., {Sip{\'{o}}cz}, B.~M., {et~al.}
  2018, \aj, 156, 123

\bibitem[{{Belokurov} {et~al.}(2018){Belokurov}, {Erkal}, {Evans}, {Koposov},
  \& {Deason}}]{Belokurov2018}
{Belokurov}, V., {Erkal}, D., {Evans}, N.~W., {Koposov}, S.~E., \& {Deason},
  A.~J. 2018, \mnras, 478, 611

\bibitem[{{Binney} \& {Tremaine}(2008)}]{Binney2008}
{Binney}, J., \& {Tremaine}, S. 2008, {Galactic Dynamics: Second Edition}

\bibitem[{{Bland-Hawthorn} \& {Gerhard}(2016)}]{BlandHawthorn2016}
{Bland-Hawthorn}, J., \& {Gerhard}, O. 2016, Annual Review of Astronomy and
  Astrophysics, 54, 529

\bibitem[{{Bond} {et~al.}(2010){Bond}, {Ivezi{\'c}}, {Sesar}, {Juri{\'c}},
  {Munn}, {Kowalski}, {Loebman}, {Ro{\v{s}}kar}, {Beers}, {Dalcanton},
  {Rockosi}, {Yanny}, {Newberg}, {Allende Prieto}, {Wilhelm}, {Lee},
  {Sivarani}, {Majewski}, {Norris}, {Bailer-Jones}, {Re Fiorentin}, {Schlegel},
  {Uomoto}, {Lupton}, {Knapp}, {Gunn}, {Covey}, {Allyn Smith}, {Miknaitis},
  {Doi}, {Tanaka}, {Fukugita}, {Kent}, {Finkbeiner}, {Quinn}, {Hawley},
  {Anderson}, {Kiuchi}, {Chen}, {Bushong}, {Sohi}, {Haggard}, {Kimball},
  {McGurk}, {Barentine}, {Brewington}, {Harvanek}, {Kleinman}, {Krzesinski},
  {Long}, {Nitta}, {Snedden}, {Lee}, {Pier}, {Harris}, {Brinkmann}, \&
  {Schneider}}]{Bond2010}
{Bond}, N.~A., {Ivezi{\'c}}, {\v{Z}}., {Sesar}, B., {et~al.} 2010, \apj, 716, 1

\bibitem[{{Bovy}(2015)}]{Bovy2015}
{Bovy}, J. 2015, \apjs, 216, 29

\bibitem[{{Bowden} {et~al.}(2015){Bowden}, {Belokurov}, \&
  {Evans}}]{Bowden2015}
{Bowden}, A., {Belokurov}, V., \& {Evans}, N.~W. 2015, \mnras, 449, 1391

\bibitem[{{Boylan-Kolchin} {et~al.}(2010){Boylan-Kolchin}, {Springel}, {White},
  \& {Jenkins}}]{BoylanKolchin2010}
{Boylan-Kolchin}, M., {Springel}, V., {White}, S.~D.~M., \& {Jenkins}, A. 2010,
  \mnras, 406, 896

\bibitem[{{Bryan} \& {Norman}(1998)}]{Bryan1998}
{Bryan}, G.~L., \& {Norman}, M.~L. 1998, \apj, 495, 80

\bibitem[{{Casetti-Dinescu} {et~al.}(2013){Casetti-Dinescu}, {Girard},
  {J{\'{\i}}lkov{\'a}}, {van Altena}, {Podest{\'a}}, \&
  {L{\'o}pez}}]{CasettiDinescu2013}
{Casetti-Dinescu}, D.~I., {Girard}, T.~M., {J{\'{\i}}lkov{\'a}}, L., {et~al.}
  2013, \aj, 146, 33

\bibitem[{{Conselice}(2014)}]{Conselice2014}
{Conselice}, C.~J. 2014, Annual Review of Astronomy and Astrophysics, 52, 291

\bibitem[{{Cunningham} {et~al.}(2016){Cunningham}, {Deason}, {Guhathakurta},
  {Rockosi}, {van der Marel}, {Toloba}, {Gilbert}, {Sohn}, \&
  {Dorman}}]{Cunningham2016}
{Cunningham}, E.~C., {Deason}, A.~J., {Guhathakurta}, P., {et~al.} 2016, \apj,
  820, doi:10.3847/0004-637X/820/1/18

\bibitem[{{Deason} {et~al.}(2012){Deason}, {Belokurov}, {Evans}, \&
  {An}}]{Deason2012}
{Deason}, A.~J., {Belokurov}, V., {Evans}, N.~W., \& {An}, J. 2012, \mnras,
  424, L44

\bibitem[{{Deason} {et~al.}(2013){Deason}, {Belokurov}, {Evans}, \&
  {Johnston}}]{Deason2013}
{Deason}, A.~J., {Belokurov}, V., {Evans}, N.~W., \& {Johnston}, K.~V. 2013,
  \apj, 763, 113

\bibitem[{{Diemand} {et~al.}(2007){Diemand}, {Kuhlen}, \&
  {Madau}}]{diemand2007}
{Diemand}, J., {Kuhlen}, M., \& {Madau}, P. 2007, \apj, 667, 859

\bibitem[{{Dotter} {et~al.}(2010){Dotter}, {Sarajedini}, {Anderson},
  {Aparicio}, {Bedin}, {Chaboyer}, {Majewski}, {Mar{\'\i}n-Franch}, {Milone},
  {Paust}, {Piotto}, {Reid}, {Rosenberg}, \& {Siegel}}]{Dotter2010}
{Dotter}, A., {Sarajedini}, A., {Anderson}, J., {et~al.} 2010, \apj, 708, 698

\bibitem[{{Eadie} \& {Juri{\'c}}(2018)}]{Eadie2018}
{Eadie}, G., \& {Juri{\'c}}, M. 2018, arXiv e-prints, arXiv:1810.10036

\bibitem[{{Eadie} {et~al.}(2015){Eadie}, {Harris}, \& {Widrow}}]{Eadie2015}
{Eadie}, G.~M., {Harris}, W.~E., \& {Widrow}, L.~M. 2015, \apj, 806, 54

\bibitem[{{Eadie} {et~al.}(2017){Eadie}, {Springford}, \& {Harris}}]{Eadie2017}
{Eadie}, G.~M., {Springford}, A., \& {Harris}, W.~E. 2017, \apj, 838, 76

\bibitem[{{Fardal} {et~al.}(2019){Fardal}, {van der Marel}, {Law}, {Sohn},
  {Sesar}, {Hernitschek}, \& {Rix}}]{Fardal2019}
{Fardal}, M.~A., {van der Marel}, R.~P., {Law}, D.~R., {et~al.} 2019, \mnras,
  483, 4724

\bibitem[{{Foreman-Mackey} {et~al.}(2013){Foreman-Mackey}, {Hogg}, {Lang}, \&
  {Goodman}}]{ForemanMackey2013}
{Foreman-Mackey}, D., {Hogg}, D.~W., {Lang}, D., \& {Goodman}, J. 2013,
  Publications of the Astronomical Society of the Pacific, 125, 306

\bibitem[{{Freeman} \& {Bland-Hawthorn}(2002)}]{Freeman2002}
{Freeman}, K., \& {Bland-Hawthorn}, J. 2002, Annual Review of Astronomy and
  Astrophysics, 40, 487

\bibitem[{{Gaia Collaboration} {et~al.}(2016{\natexlab{a}}){Gaia
  Collaboration}, {Prusti}, {de Bruijne}, {Brown}, {Vallenari}, {Babusiaux},
  {Bailer-Jones}, {Bastian}, {Biermann}, {Evans}, \& et~al.}]{Gaia}
{Gaia Collaboration}, {Prusti}, T., {de Bruijne}, J.~H.~J., {et~al.}
  2016{\natexlab{a}}, \aap, 595, A1

\bibitem[{{Gaia Collaboration} {et~al.}(2016{\natexlab{b}}){Gaia
  Collaboration}, {Brown}, {Vallenari}, {Prusti}, {de Bruijne}, {Mignard},
  {Drimmel}, {Babusiaux}, {Bailer-Jones}, {Bastian}, \& et~al.}]{GaiaDR1}
{Gaia Collaboration}, {Brown}, A.~G.~A., {Vallenari}, A., {et~al.}
  2016{\natexlab{b}}, \aap, 595, A2

\bibitem[{{Gaia Collaboration} {et~al.}(2018{\natexlab{a}}){Gaia
  Collaboration}, {Brown}, {Vallenari}, {Prusti}, {de Bruijne}, {Babusiaux},
  {Bailer-Jones}, {Biermann}, {Evans}, {Eyer}, {Jansen}, {Jordi}, {Klioner},
  {Lammers}, {Lindegren}, {Luri}, {Mignard}, {Panem}, {Pourbaix}, {Randich},
  {Sartoretti}, {Siddiqui}, {Soubiran}, {van Leeuwen}, {Walton}, {Arenou},
  {Bastian}, {Cropper}, {Drimmel}, {Katz}, {Lattanzi}, {Bakker}, {Cacciari},
  {Casta{\~n}eda}, {Chaoul}, {Cheek}, {De Angeli}, {Fabricius}, {Guerra},
  {Holl}, {Masana}, {Messineo}, {Mowlavi}, {Nienartowicz}, {Panuzzo},
  {Portell}, {Riello}, {Seabroke}, {Tanga}, {Th{\'e}venin}, {Gracia-Abril},
  {Comoretto}, {Garcia-Reinaldos}, {Teyssier}, {Altmann}, {Andrae}, {Audard},
  {Bellas-Velidis}, {Benson}, {Berthier}, {Blomme}, {Burgess}, {Busso},
  {Carry}, {Cellino}, {Clementini}, {Clotet}, {Creevey}, {Davidson}, {De
  Ridder}, {Delchambre}, {Dell'Oro}, {Ducourant}, {Fern{\'a}ndez-
  Hern{\'a}ndez}, {Fouesneau}, {Fr{\'e}mat}, {Galluccio}, {Garc{\'\i}a-Torres},
  {Gonz{\'a}lez-N{\'u}{\~n}ez}, {Gonz{\'a}lez-Vidal}, {Gosset}, {Guy},
  {Halbwachs}, {Hambly}, {Harrison}, {Hern{\'a}ndez}, {Hestroffer}, {Hodgkin},
  {Hutton}, {Jasniewicz}, {Jean-Antoine-Piccolo}, {Jordan}, {Korn},
  {Krone-Martins}, {Lanzafame}, {Lebzelter}, {L{\"o}ffler}, {Manteiga},
  {Marrese}, {Mart{\'\i}n-Fleitas}, {Moitinho}, {Mora}, {Muinonen}, {Osinde},
  {Pancino}, {Pauwels}, {Petit}, {Recio-Blanco}, {Richards}, {Rimoldini},
  {Robin}, {Sarro}, {Siopis}, {Smith}, {Sozzetti}, {S{\"u}veges}, {Torra}, {van
  Reeven}, {Abbas}, {Abreu Aramburu}, {Accart}, {Aerts}, {Altavilla},
  {{\'A}lvarez}, {Alvarez}, {Alves}, {Anderson}, {Andrei}, {Anglada Varela},
  {Antiche}, {Antoja}, {Arcay}, {Astraatmadja}, {Bach}, {Baker},
  {Balaguer-N{\'u}{\~n}ez}, {Balm}, {Barache}, {Barata}, {Barbato}, {Barblan},
  {Barklem}, {Barrado}, {Barros}, {Barstow}, {Bartholom{\'e} Mu{\~n}oz},
  {Bassilana}, {Becciani}, {Bellazzini}, {Berihuete}, {Bertone}, {Bianchi},
  {Bienaym{\'e}}, {Blanco-Cuaresma}, {Boch}, {Boeche}, {Bombrun}, {Borrachero},
  {Bossini}, {Bouquillon}, {Bourda}, {Bragaglia}, {Bramante}, {Breddels},
  {Bressan}, {Brouillet}, {Br{\"u}semeister}, {Brugaletta}, {Bucciarelli},
  {Burlacu}, {Busonero}, {Butkevich}, {Buzzi}, {Caffau}, {Cancelliere},
  {Cannizzaro}, {Cantat-Gaudin}, {Carballo}, {Carlucci}, {Carrasco},
  {Casamiquela}, {Castellani}, {Castro-Ginard}, {Charlot}, {Chemin},
  {Chiavassa}, {Cocozza}, {Costigan}, {Cowell}, {Crifo}, {Crosta}, {Crowley},
  {Cuypers}, {Dafonte}, {Damerdji}, {Dapergolas}, {David}, {David}, {de
  Laverny}, {De Luise}, {De March}, {de Martino}, {de Souza}, {de Torres},
  {Debosscher}, {del Pozo}, {Delbo}, {Delgado}, {Delgado}, {Di Matteo},
  {Diakite}, {Diener}, {Distefano}, {Dolding}, {Drazinos}, {Dur{\'a}n},
  {Edvardsson}, {Enke}, {Eriksson}, {Esquej}, {Eynard Bontemps}, {Fabre},
  {Fabrizio}, {Faigler}, {Falc{\~a}o}, {Farr{\`a}s Casas}, {Federici},
  {Fedorets}, {Fernique}, {Figueras}, {Filippi}, {Findeisen}, {Fonti},
  {Fraile}, {Fraser}, {Fr{\'e}zouls}, {Gai}, {Galleti}, {Garabato},
  {Garc{\'\i}a-Sedano}, {Garofalo}, {Garralda}, {Gavel}, {Gavras}, {Gerssen},
  {Geyer}, {Giacobbe}, {Gilmore}, {Girona}, {Giuffrida}, {Glass}, {Gomes},
  {Granvik}, {Gueguen}, {Guerrier}, {Guiraud}, {Guti{\'e}rrez-S{\'a}nchez},
  {Haigron}, {Hatzidimitriou}, {Hauser}, {Haywood}, {Heiter}, {Helmi}, {Heu},
  {Hilger}, {Hobbs}, {Hofmann}, {Holland}, {Huckle}, {Hypki}, {Icardi},
  {Jan{\ss}en}, {Jevardat de Fombelle}, {Jonker}, {Juh{\'a}sz}, {Julbe},
  {Karampelas}, {Kewley}, {Klar}, {Kochoska}, {Kohley}, {Kolenberg},
  {Kontizas}, {Kontizas}, {Koposov}, {Kordopatis}, {Kostrzewa-Rutkowska},
  {Koubsky}, {Lambert}, {Lanza}, {Lasne}, {Lavigne}, {Le Fustec}, {Le
  Poncin-Lafitte}, {Lebreton}, {Leccia}, {Leclerc}, {Lecoeur-Taibi},
  {Lenhardt}, {Leroux}, {Liao}, {Licata}, {Lindstr{\o}m}, {Lister}, {Livanou},
  {Lobel}, {L{\'o}pez}, {Managau}, {Mann}, {Mantelet}, {Marchal}, {Marchant},
  {Marconi}, {Marinoni}, {Marschalk{\'o}}, {Marshall}, {Martino}, {Marton},
  {Mary}, {Massari}, {Matijevi{\v{c}}}, {Mazeh}, {McMillan}, {Messina},
  {Michalik}, {Millar}, {Molina}, {Molinaro}, {Moln{\'a}r}, {Montegriffo},
  {Mor}, {Morbidelli}, {Morel}, {Morris}, {Mulone}, {Muraveva}, {Musella},
  {Nelemans}, {Nicastro}, {Noval}, {O'Mullane}, {Ord{\'e}novic},
  {Ord{\'o}{\~n}ez-Blanco}, {Osborne}, {Pagani}, {Pagano}, {Pailler},
  {Palacin}, {Palaversa}, {Panahi}, {Pawlak}, {Piersimoni}, {Pineau}, {Plachy},
  {Plum}, {Poggio}, {Poujoulet}, {Pr{\v{s}}a}, {Pulone}, {Racero}, {Ragaini},
  {Rambaux}, {Ramos-Lerate}, {Regibo}, {Reyl{\'e}}, {Riclet}, {Ripepi}, {Riva},
  {Rivard}, {Rixon}, {Roegiers}, {Roelens}, {Romero-G{\'o}mez}, {Rowell},
  {Royer}, {Ruiz-Dern}, {Sadowski}, {Sagrist{\`a} Sell{\'e}s}, {Sahlmann},
  {Salgado}, {Salguero}, {Sanna}, {Santana- Ros}, {Sarasso}, {Savietto},
  {Schultheis}, {Sciacca}, {Segol}, {Segovia}, {S{\'e}gransan}, {Shih},
  {Siltala}, {Silva}, {Smart}, {Smith}, {Solano}, {Solitro}, {Sordo}, {Soria
  Nieto}, {Souchay}, {Spagna}, {Spoto}, {Stampa}, {Steele},
  {Steidelm{\"u}ller}, {Stephenson}, {Stoev}, {Suess}, {Surdej}, {Szabados},
  {Szegedi-Elek}, {Tapiador}, {Taris}, {Tauran}, {Taylor}, {Teixeira},
  {Terrett}, {Teyssandier}, {Thuillot}, {Titarenko}, {Torra Clotet}, {Turon},
  {Ulla}, {Utrilla}, {Uzzi}, {Vaillant}, {Valentini}, {Valette}, {van Elteren},
  {Van Hemelryck}, {van Leeuwen}, {Vaschetto}, {Vecchiato}, {Veljanoski},
  {Viala}, {Vicente}, {Vogt}, {von Essen}, {Voss}, {Votruba}, {Voutsinas},
  {Walmsley}, {Weiler}, {Wertz}, {Wevers}, {Wyrzykowski}, {Yoldas},
  {{\v{Z}}erjal}, {Ziaeepour}, {Zorec}, {Zschocke}, {Zucker}, {Zurbach}, \&
  {Zwitter}}]{GaiaDR2}
---. 2018{\natexlab{a}}, \aap, 616, A1

\bibitem[{{Gaia Collaboration} {et~al.}(2018{\natexlab{b}}){Gaia
  Collaboration}, {Helmi}, {van Leeuwen}, {McMillan}, {Massari}, {Antoja},
  {Robin}, {Lindegren}, {Bastian}, {Arenou}, {Babusiaux}, {Biermann},
  {Breddels}, {Hobbs}, {Jordi}, {Pancino}, {Reyl{\'e}}, {Veljanoski}, {Brown},
  {Vallenari}, {Prusti}, {de Bruijne}, {Bailer-Jones}, {Evans}, {Eyer},
  {Jansen}, {Klioner}, {Lammers}, {Luri}, {Mignard}, {Panem}, {Pourbaix},
  {Randich}, {Sartoretti}, {Siddiqui}, {Soubiran}, {Walton}, {Cropper},
  {Drimmel}, {Katz}, {Lattanzi}, {Bakker}, {Cacciari}, {Casta{\~n}eda},
  {Chaoul}, {Cheek}, {De Angeli}, {Fabricius}, {Guerra}, {Holl}, {Masana},
  {Messineo}, {Mowlavi}, {Nienartowicz}, {Panuzzo}, {Portell}, {Riello},
  {Seabroke}, {Tanga}, {Th{\'e}venin}, {Gracia-Abril}, {Comoretto},
  {Garcia-Reinaldos}, {Teyssier}, {Altmann}, {Andrae}, {Audard},
  {Bellas-Velidis}, {Benson}, {Berthier}, {Blomme}, {Burgess}, {Busso},
  {Carry}, {Cellino}, {Clementini}, {Clotet}, {Creevey}, {Davidson}, {De
  Ridder}, {Delchambre}, {Dell'Oro}, {Ducourant},
  {Fern{\'a}ndez-Hern{\'a}ndez}, {Fouesneau}, {Fr{\'e}mat}, {Galluccio},
  {Garc{\'\i}a-Torres}, {Gonz{\'a}lez-N{\'u}{\~n}ez}, {Gonz{\'a}lez- Vidal},
  {Gosset}, {Guy}, {Halbwachs}, {Hambly}, {Harrison}, {Hern{\'a}ndez},
  {Hestroffer}, {Hodgkin}, {Hutton}, {Jasniewicz}, {Jean-Antoine- Piccolo},
  {Jordan}, {Korn}, {Krone- Martins}, {Lanzafame}, {Lebzelter}, {L{\"o}ffler},
  {Manteiga}, {Marrese}, {Mart{\'\i}n-Fleitas}, {Moitinho}, {Mora}, {Muinonen},
  {Osinde}, {Pauwels}, {Petit}, {Recio-Blanco}, {Richards}, {Rimoldini},
  {Sarro}, {Siopis}, {Smith}, {Sozzetti}, {S{\"u}veges}, {Torra}, {van Reeven},
  {Abbas}, {Abreu Aramburu}, {Accart}, {Aerts}, {Altavilla}, {{\'A}lvarez},
  {Alvarez}, {Alves}, {Anderson}, {Andrei}, {Anglada Varela}, {Antiche},
  {Arcay}, {Astraatmadja}, {Bach}, {Baker}, {Balaguer-N{\'u}{\~n}ez}, {Balm},
  {Barache}, {Barata}, {Barbato}, {Barblan}, {Barklem}, {Barrado}, {Barros},
  {Barstow}, {Bartholom{\'e} Mu{\~n}oz}, {Bassilana}, {Becciani}, {Bellazzini},
  {Berihuete}, {Bertone}, {Bianchi}, {Bienaym{\'e}}, {Blanco-Cuaresma}, {Boch},
  {Boeche}, {Bombrun}, {Borrachero}, {Bossini}, {Bouquillon}, {Bourda},
  {Bragaglia}, {Bramante}, {Bressan}, {Brouillet}, {Br{\"u}semeister},
  {Brugaletta}, {Bucciarelli}, {Burlacu}, {Busonero}, {Butkevich}, {Buzzi},
  {Caffau}, {Cancelliere}, {Cannizzaro}, {Cantat-Gaudin}, {Carballo},
  {Carlucci}, {Carrasco}, {Casamiquela}, {Castellani}, {Castro-Ginard},
  {Charlot}, {Chemin}, {Chiavassa}, {Cocozza}, {Costigan}, {Cowell}, {Crifo},
  {Crosta}, {Crowley}, {Cuypers}, {Dafonte}, {Damerdji}, {Dapergolas}, {David},
  {David}, {de Laverny}, {De Luise}, {De March}, {de Martino}, {de Souza}, {de
  Torres}, {Debosscher}, {del Pozo}, {Delbo}, {Delgado}, {Delgado}, {Di
  Matteo}, {Diakite}, {Diener}, {Distefano}, {Dolding}, {Drazinos},
  {Dur{\'a}n}, {Edvardsson}, {Enke}, {Eriksson}, {Esquej}, {Eynard Bontemps},
  {Fabre}, {Fabrizio}, {Faigler}, {Falc{\~a}o}, {Farr{\`a}s Casas}, {Federici},
  {Fedorets}, {Fernique}, {Figueras}, {Filippi}, {Findeisen}, {Fonti},
  {Fraile}, {Fraser}, {Fr{\'e}zouls}, {Gai}, {Galleti}, {Garabato},
  {Garc{\'\i}a-Sedano}, {Garofalo}, {Garralda}, {Gavel}, {Gavras}, {Gerssen},
  {Geyer}, {Giacobbe}, {Gilmore}, {Girona}, {Giuffrida}, {Glass}, {Gomes},
  {Granvik}, {Gueguen}, {Guerrier}, {Guiraud}, {Guti{\'e}rrez-S{\'a}nchez},
  {Hofmann}, {Holland}, {Huckle}, {Hypki}, {Icardi}, {Jan{\ss}en}, {Jevardat de
  Fombelle}, {Jonker}, {Juh{\'a}sz}, {Julbe}, {Karampelas}, {Kewley}, {Klar},
  {Kochoska}, {Kohley}, {Kolenberg}, {Kontizas}, {Kontizas}, {Koposov},
  {Kordopatis}, {Kostrzewa-Rutkowska}, {Koubsky}, {Lambert}, {Lanza}, {Lasne},
  {Lavigne}, {Le Fustec}, {Le Poncin- Lafitte}, {Lebreton}, {Leccia},
  {Leclerc}, {Lecoeur-Taibi}, {Lenhardt}, {Leroux}, {Liao}, {Licata},
  {Lindstr{\o}m}, {Lister}, {Livanou}, {Lobel}, {L{\'o}pez}, {Managau}, {Mann},
  {Mantelet}, {Marchal}, {Marchant}, {Marconi}, {Marinoni}, {Marschalk{\'o}},
  {Marshall}, {Martino}, {Marton}, {Mary}, {Matijevi{\v{c}}}, {Mazeh},
  {Messina}, {Michalik}, {Millar}, {Molina}, {Molinaro}, {Moln{\'a}r},
  {Montegriffo}, {Mor}, {Morbidelli}, {Morel}, {Morris}, {Mulone}, {Muraveva},
  {Musella}, {Nelemans}, {Nicastro}, {Noval}, {O'Mullane}, {Ord{\'e}novic},
  {Ord{\'o}{\~n}ez- Blanco}, {Osborne}, {Pagani}, {Pagano}, {Pailler},
  {Palacin}, {Palaversa}, {Panahi}, {Pawlak}, {Piersimoni}, {Pineau}, {Plachy},
  {Plum}, {Poggio}, {Poujoulet}, {Pr{\v{s}}a}, {Pulone}, {Racero}, {Ragaini},
  {Rambaux}, {Ramos- Lerate}, {Regibo}, {Riclet}, {Ripepi}, {Riva}, {Rivard},
  {Rixon}, {Roegiers}, {Roelens}, {Romero-G{\'o}mez}, {Rowell}, {Royer},
  {Ruiz-Dern}, {Sadowski}, {Sagrist{\`a} Sell{\'e}s}, {Sahlmann}, {Salgado},
  {Salguero}, {Sanna}, {Santana-Ros}, {Sarasso}, {Savietto}, {Schultheis},
  {Sciacca}, {Segol}, {Segovia}, {S{\'e}gransan}, {Shih}, {Siltala}, {Silva},
  {Smart}, {Smith}, {Solano}, {Solitro}, {Sordo}, {Soria Nieto}, {Souchay},
  {Spagna}, {Spoto}, {Stampa}, {Steele}, {Steidelm{\"u}ller}, {Stephenson},
  {Stoev}, {Suess}, {Surdej}, {Szabados}, {Szegedi-Elek}, {Tapiador}, {Taris},
  {Tauran}, {Taylor}, {Teixeira}, {Terrett}, {Teyssandier}, {Thuillot},
  {Titarenko}, {Torra Clotet}, {Turon}, {Ulla}, {Utrilla}, {Uzzi}, {Vaillant},
  {Valentini}, {Valette}, {van Elteren}, {Van Hemelryck}, {van Leeuwen},
  {Vaschetto}, {Vecchiato}, {Viala}, {Vicente}, {Vogt}, {von Essen}, {Voss},
  {Votruba}, {Voutsinas}, {Walmsley}, {Weiler}, {Wertz}, {Wevems},
  {Wyrzykowski}, {Yoldas}, {{\v{Z}}erjal}, {Ziaeepour}, {Zorec}, {Zschocke},
  {Zucker}, {Zurbach}, \& {Zwitter}}]{GaiaGCs}
{Gaia Collaboration}, {Helmi}, A., {van Leeuwen}, F., {et~al.}
  2018{\natexlab{b}}, \aap, 616, A12

\bibitem[{{Gibbons} {et~al.}(2014){Gibbons}, {Belokurov}, \&
  {Evans}}]{Gibbons2014}
{Gibbons}, S.~L.~J., {Belokurov}, V., \& {Evans}, N.~W. 2014, \mnras, 445, 3788

\bibitem[{{Harris}(1996)}]{Harris1996}
{Harris}, W.~E. 1996, \aj, 112, 1487

\bibitem[{{Harris}(2001)}]{Harris2001}
---. 2001, {Globular Cluster Systems}, 223

\bibitem[{{Helmi}(2008)}]{Helmi2008}
{Helmi}, A. 2008, Astronomy and Astrophysics Review, 15, 145

\bibitem[{{H{\o}g} {et~al.}(2000){H{\o}g}, {Fabricius}, {Makarov}, {Urban},
  {Corbin}, {Wycoff}, {Bastian}, {Schwekendiek}, \& {Wicenec}}]{Hog2000}
{H{\o}g}, E., {Fabricius}, C., {Makarov}, V.~V., {et~al.} 2000, \aap, 355, L27

\bibitem[{{Ivezi{\'c}} {et~al.}(2012){Ivezi{\'c}}, {Beers}, \&
  {Juri{\'c}}}]{Ivezic2012}
{Ivezi{\'c}}, {\v{Z}}., {Beers}, T.~C., \& {Juri{\'c}}, M. 2012, Annual Review
  of Astronomy and Astrophysics, 50, 251

\bibitem[{{Kafle} {et~al.}(2012){Kafle}, {Sharma}, {Lewis}, \&
  {Bland-Hawthorn}}]{Kafle2012}
{Kafle}, P.~R., {Sharma}, S., {Lewis}, G.~F., \& {Bland-Hawthorn}, J. 2012,
  \apj, 761, doi:10.1088/0004-637X/761/2/98

\bibitem[{{Kalirai} {et~al.}(2007){Kalirai}, {Anderson}, {Richer}, {King},
  {Brewer}, {Carraro}, {Davis}, {Fahlman}, {Hansen}, {Hurley}, {L{\'e}pine},
  {Reitzel}, {Rich}, {Shara}, \& {Stetson}}]{Kalirai2007}
{Kalirai}, J.~S., {Anderson}, J., {Richer}, H.~B., {et~al.} 2007, \apj, 657,
  L93

\bibitem[{{Kallivayalil} {et~al.}(2013){Kallivayalil}, {van der Marel},
  {Besla}, {Anderson}, \& {Alcock}}]{Kallivayalil2013}
{Kallivayalil}, N., {van der Marel}, R.~P., {Besla}, G., {Anderson}, J., \&
  {Alcock}, C. 2013, \apj, 764, doi:10.1088/0004-637X/764/2/161

\bibitem[{{Klypin} {et~al.}(2002){Klypin}, {Zhao}, \&
  {Somerville}}]{Klypin2002}
{Klypin}, A., {Zhao}, H., \& {Somerville}, R.~S. 2002, \apj, 573, 597

\bibitem[{{Klypin} {et~al.}(2011){Klypin}, {Trujillo-Gomez}, \&
  {Primack}}]{Klypin2011}
{Klypin}, A.~A., {Trujillo-Gomez}, S., \& {Primack}, J. 2011, \apj, 740,
  doi:10.1088/0004-637X/740/2/102

\bibitem[{{Koposov} {et~al.}(2010){Koposov}, {Rix}, \& {Hogg}}]{Koposov2010}
{Koposov}, S.~E., {Rix}, H.-W., \& {Hogg}, D.~W. 2010, \apj, 712, 260

\bibitem[{{K{\"u}pper} {et~al.}(2015){K{\"u}pper}, {Balbinot}, {Bonaca},
  {Johnston}, {Hogg}, {Kroupa}, \& {Santiago}}]{Kuepper2015}
{K{\"u}pper}, A. H.~W., {Balbinot}, E., {Bonaca}, A., {et~al.} 2015, \apj, 803,
  doi:10.1088/0004-637X/803/2/80

\bibitem[{{Li} \& {White}(2008)}]{Li2008}
{Li}, Y.-S., \& {White}, S. D.~M. 2008, \mnras, 384, 1459

\bibitem[{{Massari} {et~al.}(2017){Massari}, {Posti}, {Helmi}, {Fiorentino}, \&
  {Tolstoy}}]{Massari2017}
{Massari}, D., {Posti}, L., {Helmi}, A., {Fiorentino}, G., \& {Tolstoy}, E.
  2017, \aap, 598, L9

\bibitem[{{McMillan}(2011)}]{McMillan2011}
{McMillan}, P.~J. 2011, \mnras, 414, 2446

\bibitem[{{Michalik} {et~al.}(2015){Michalik}, {Lindegren}, \&
  {Hobbs}}]{Michalik2015}
{Michalik}, D., {Lindegren}, L., \& {Hobbs}, D. 2015, \aap, 574, A115

\bibitem[{{Myeong} {et~al.}(2018{\natexlab{a}}){Myeong}, {Evans}, {Belokurov},
  {Sanders}, \& {Koposov}}]{Myeong2018}
{Myeong}, G.~C., {Evans}, N.~W., {Belokurov}, V., {Sanders}, J.~L., \&
  {Koposov}, S.~E. 2018{\natexlab{a}}, \apjl, 856, L26

\bibitem[{{Myeong} {et~al.}(2018{\natexlab{b}}){Myeong}, {Evans}, {Belokurov},
  {Sanders}, \& {Koposov}}]{Myeong2018b}
---. 2018{\natexlab{b}}, \apj, 863, L28

\bibitem[{{Navarro} {et~al.}(1996){Navarro}, {Frenk}, \& {White}}]{Navarro1996}
{Navarro}, J.~F., {Frenk}, C.~S., \& {White}, S.~D.~M. 1996, \apj, 462, 563

\bibitem[{{Patel} {et~al.}(2018){Patel}, {Besla}, {Mandel}, \& {Sohn}}]{Pa18}
{Patel}, E., {Besla}, G., {Mandel}, K., \& {Sohn}, S.~T. 2018, \apj, 857, 78

\bibitem[{{Piatek} {et~al.}(2016){Piatek}, {Pryor}, \&
  {Olszewski}}]{Piatek2016}
{Piatek}, S., {Pryor}, C., \& {Olszewski}, E.~W. 2016, \aj, 152,
  doi:10.3847/0004-6256/152/6/166

\bibitem[{{Posti} \& {Helmi}(2019)}]{Posti2019}
{Posti}, L., \& {Helmi}, A. 2019, \aap, 621, A56

\bibitem[{{Price-Whelan}(2017)}]{PriceWhelan2017}
{Price-Whelan}, A.~M. 2017, The Journal of Open Source Software, 2, 388

\bibitem[{{Sakamoto} {et~al.}(2003){Sakamoto}, {Chiba}, \&
  {Beers}}]{Sakamoto2003}
{Sakamoto}, T., {Chiba}, M., \& {Beers}, T.~C. 2003, \aap, 397, 899

\bibitem[{{Sanderson} {et~al.}(2017){Sanderson}, {Hartke}, \& {Helmi}}]{Sa17}
{Sanderson}, R.~E., {Hartke}, J., \& {Helmi}, A. 2017, \apj, 836, 234

\bibitem[{{Sch{\"o}nrich} {et~al.}(2010){Sch{\"o}nrich}, {Binney}, \&
  {Dehnen}}]{Schoenrich2010}
{Sch{\"o}nrich}, R., {Binney}, J., \& {Dehnen}, W. 2010, \mnras, 403, 1829

\bibitem[{{Sirko} {et~al.}(2004){Sirko}, {Goodman}, {Knapp}, {Brinkmann},
  {Ivezi{\'c}}, {Knerr}, {Schlegel}, {Schneider}, \& {York}}]{Sirko2004}
{Sirko}, E., {Goodman}, J., {Knapp}, G.~R., {et~al.} 2004, \aj, 127, 914

\bibitem[{{Sohn} {et~al.}(2018){Sohn}, {Watkins}, {Fardal}, {van der Marel},
  {Deason}, {Besla}, \& {Bellini}}]{Sohn2018}
{Sohn}, S.~T., {Watkins}, L.~L., {Fardal}, M.~A., {et~al.} 2018, \apj, 862,
  doi:10.3847/1538-4357/aacd0b

\bibitem[{{Sohn} {et~al.}(2017){Sohn}, {Patel}, {Besla}, {van der Marel},
  {Bullock}, {Strigari}, {van de Ven}, {Walker}, \& {Bellini}}]{Sohn2017}
{Sohn}, S.~T., {Patel}, E., {Besla}, G., {et~al.} 2017, \apj, 849,
  doi:10.3847/1538-4357/aa917b

\bibitem[{{van der Marel} {et~al.}(2012{\natexlab{a}}){van der Marel}, {Besla},
  {Cox}, {Sohn}, \& {Anderson}}]{vanderMarel2012b}
{van der Marel}, R.~P., {Besla}, G., {Cox}, T.~J., {Sohn}, S.~T., \&
  {Anderson}, J. 2012{\natexlab{a}}, \apj, 753, doi:10.1088/0004-637X/753/1/9

\bibitem[{{van der Marel} {et~al.}(2012{\natexlab{b}}){van der Marel},
  {Fardal}, {Besla}, {Beaton}, {Sohn}, {Anderson}, {Brown}, \&
  {Guhathakurta}}]{vanderMarel2012a}
{van der Marel}, R.~P., {Fardal}, M., {Besla}, G., {et~al.} 2012{\natexlab{b}},
  \apj, 753, doi:10.1088/0004-637X/753/1/8

\bibitem[{{Vasiliev}(2019)}]{Vasiliev2019}
{Vasiliev}, E. 2019, \mnras, 176

\bibitem[{{Wang} {et~al.}(2015){Wang}, {Han}, {Cooper}, {Cole}, {Frenk}, \&
  {Lowing}}]{Wang2015}
{Wang}, W., {Han}, J., {Cooper}, A.~P., {et~al.} 2015, \mnras, 453, 377

\bibitem[{{Watkins} {et~al.}(2010){Watkins}, {Evans}, \& {An}}]{Watkins2010}
{Watkins}, L.~L., {Evans}, N.~W., \& {An}, J.~H. 2010, \mnras, 406, 264

\bibitem[{{Watkins} \& {van der Marel}(2017)}]{Watkins2017}
{Watkins}, L.~L., \& {van der Marel}, R.~P. 2017, \apj, 839, 89

\bibitem[{{Watkins} {et~al.}(2015){Watkins}, {van der Marel}, {Bellini}, \&
  {Anderson}}]{Watkins2015}
{Watkins}, L.~L., {van der Marel}, R.~P., {Bellini}, A., \& {Anderson}, J.
  2015, \apj, 812, doi:10.1088/0004-637X/812/2/149

\bibitem[{{Wilkinson} \& {Evans}(1999)}]{Wilkinson1999}
{Wilkinson}, M.~I., \& {Evans}, N.~W. 1999, \mnras, 310, 645

\bibitem[{{Zinn}(1993)}]{Zinn1993}
{Zinn}, R. 1993, in The globular clusters-galaxy connection. Astronomical
  Society of the Pacific Conference Series, Volume 48, Proceedings of the 11th
  Santa Cruz Summer Workshop in Astronomy and Astrophysics, held July 19-29,
  1992, at the University of California, Santa Cruz, San Francisco:
  Astronomical Society of the Pacific (ASP), |c1993, edited by Graeme H. Smith,
  and Jean P. Brodie, ISBN 0-937707-67-8. LC QB856 .S26 1992, p.38, Vol.~48, 38

\end{thebibliography}

\appendix

\section{Galactocentric positions and motions}
\label{appendix:galactocentric}

In \autoref{section:data}, we calculated Galactocentric positions and velocities for all 75 GCs in \citet{GaiaGCs}. We provide these positions and motions in both spherical and Cartesian coordinates in \autoref{table:galactocentric} along with their uncertainties and the correlations between velocity components. Note that the astrometric measurements and heliocentric Cartesian positions and velocities are provided in \citet{GaiaGCs}.\footnote{Available electronically at \url{https://www.astro.rug.nl/~ahelmi/research/dr2-dggc/}}

This table will eventually be available online through the journal, however in the meantime, the data is available upon request.

\begin{splitdeluxetable*}{ccccccccccBcccccccccc}
\tablecaption{Galactocentric positions and velocities in spherical and Cartesian coordinates for the full \textit{Gaia} GC sample. \label{table:galactocentric}} 
\tablehead{
   \colhead{Name} & 
   \colhead{$r$} & 
   \colhead{$\theta$} & 
   \colhead{$\phi$} &
   \colhead{$v_\mathrm{r}$} & 
   \colhead{$v_\theta$} &
   \colhead{$v_\phi$} &
   \colhead{$C_{\mathrm{r}\theta}$} & 
   \colhead{$C_{\mathrm{r}\phi}$} &
   \colhead{$C_{\theta\phi}$} &
   \colhead{$x$} & 
   \colhead{$y$} & 
   \colhead{$z$} &
   \colhead{$v_\mathrm{x}$} & 
   \colhead{$v_\mathrm{y}$} &
   \colhead{$v_\mathrm{z}$} &
   \colhead{$C_\mathrm{xy}$} & 
   \colhead{$C_\mathrm{xz}$} &
   \colhead{$C_\mathrm{yz}$} &
   \colhead{$v$} \\
   \colhead{} & 
   \colhead{(kpc)} & 
   \colhead{(deg)} & 
   \colhead{(deg)} & 
   \colhead{(\kms)} & 
   \colhead{(\kms)} & 
   \colhead{(\kms)} &
   \colhead{} & 
   \colhead{} & 
   \colhead{} & 
   \colhead{(kpc)} & 
   \colhead{(kpc)} & 
   \colhead{(kpc)} & 
   \colhead{(\kms)} & 
   \colhead{(\kms)} & 
   \colhead{(\kms)} &
   \colhead{} & 
   \colhead{} & 
   \colhead{} & 
   \colhead{(\kms)}
}
\startdata
NGC\,0104 &  7.6 $\pm$ 0.1 & -24.7 $\pm$  0.7 & 202.0 $\pm$  0.8 &  -12.0 $\pm$  2.8 &   44.8 $\pm$  1.5 & -187.9 $\pm$  5.2 & -0.006 &  0.004 & -0.001 & -6.4 $\pm$ 0.2 &  -2.6 $\pm$  0.1 &  -3.2 $\pm$  0.1 &  -77.6 $\pm$  2.3 &  171.3 $\pm$  5.9 &   45.7 $\pm$  0.9 & -0.000 &  0.000 & -0.000 &  193.6 $\pm$  5.0 \\
NGC\,0288 & 12.2 $\pm$ 0.2 & -46.8 $\pm$  0.8 & 179.7 $\pm$  0.0 &  -31.9 $\pm$  1.1 &   41.0 $\pm$  1.0 &   46.0 $\pm$  8.6 & -0.001 &  0.001 & -0.001 & -8.4 $\pm$ 0.2 &   0.0 $\pm$  0.0 &  -8.9 $\pm$  0.2 &   -8.3 $\pm$  1.3 &  -46.0 $\pm$  8.6 &   51.3 $\pm$  0.7 &  0.007 & -0.000 & -0.000 &   69.7 $\pm$  5.7 \\
NGC\,0362 &  9.5 $\pm$ 0.1 & -40.6 $\pm$  0.7 & 224.5 $\pm$  1.4 &  143.9 $\pm$  4.4 &   30.2 $\pm$  7.3 &    4.3 $\pm$  4.8 &  0.007 &  0.008 & -0.044 & -5.2 $\pm$ 0.2 &  -5.1 $\pm$  0.1 &  -6.2 $\pm$  0.1 &  -89.0 $\pm$  4.2 &  -93.5 $\pm$  7.1 &  -70.8 $\pm$  2.0 &  0.002 & -0.001 & -0.001 &  147.3 $\pm$  5.7 \\
NGC\,1851 & 16.9 $\pm$ 0.3 & -24.3 $\pm$  0.3 & 215.5 $\pm$  0.5 &  131.5 $\pm$  3.1 &  -30.3 $\pm$  3.3 &    7.0 $\pm$  4.7 & -0.001 &  0.013 & -0.039 & -12.5 $\pm$ 0.2 &  -8.9 $\pm$  0.2 &  -6.9 $\pm$  0.2 &  -83.4 $\pm$  1.3 &  -68.1 $\pm$  5.8 &  -81.6 $\pm$  2.3 & -0.000 &  0.000 & -0.001 &  135.2 $\pm$  3.0 \\
NGC\,1904 & 19.0 $\pm$ 0.3 & -19.4 $\pm$  0.2 & 207.4 $\pm$  0.4 &   43.9 $\pm$  2.7 &   20.4 $\pm$  2.6 &   -8.3 $\pm$  5.9 &  0.000 & -0.023 & -0.068 & -15.9 $\pm$ 0.2 &  -8.3 $\pm$  0.2 &  -6.3 $\pm$  0.1 &  -46.6 $\pm$  2.0 &  -14.8 $\pm$  6.3 &    4.6 $\pm$  2.4 & -0.008 & -0.016 &  0.107 &   49.6 $\pm$  2.1 \\
\enddata
\tablecomments{This table is published in its entirety in the machine-readable format. A portion is shown here for guidance regarding its form and content.}
\end{splitdeluxetable*}

\end{document}